\documentclass[aps,prl,twocolumn,reprint]{revtex4-1} 
\usepackage{amsmath,amssymb,amsthm,bm,braket,ascmac,bbm}
\usepackage[pdftex]{graphicx}
\usepackage{color}
\usepackage{comment}
\usepackage{braket}
\usepackage{mathtools}
\usepackage{docmute}
\usepackage{bm}
\usepackage{multirow}

\newcommand{\imi}{\mathrm{i}}
\newcommand{\paren}[1]{\left(#1\right)}
\newcommand{\bck}[1]{\left[#1\right]}
\newcommand{\bce}[1]{\left\{#1\right\}}

\newcommand{\wsup}{w}
\newcommand{\knm}{t}

\begin{document}
\title{Explicit Construction of Local Conserved Quantities in the XYZ Spin-1/2 Chain}

\author{Yuji Nozawa}\email{y.nozawa@issp.u-tokyo.ac.jp}
\affiliation{The Institute for Solid State Physics, The University of Tokyo, Kashiwa, Chiba 277-8581, Japan}
\author{Kouhei Fukai}
\affiliation{The Institute for Solid State Physics, The University of Tokyo, Kashiwa, Chiba 277-8581, Japan}

\date{\today}
\begin{abstract}
We present a rigorous explicit expression for an extensive number of local conserved quantities in the XYZ spin-$1/2$ chain with general coupling constants. All the coefficients of operators in each local conserved quantity are calculated.
We also confirm that our result can be applied to the case of the XXZ chain with a magnetic field in the z-axis direction.
\end{abstract}
\maketitle

\textit{Introduction.---}
Understanding and describing nonequilibrium phenomena in quantum many-body systems is one of the challenging problems in physics. 
Particularly for the past two decades, nonequilibrium phenomena in integrable systems have been attracting more attention owing to their experimental realization with ultracold atomic gases~\cite{kinoshita2006quantum,RevModPhys.80.885,RevModPhys.83.863,Langen_2016}.
An extensive number of local conserved quantities, which characterizes integrable systems, are key elements of nonequilibrium phenomena. For example, these quantities prevent systems from relaxing to a thermal state, which is described by the canonical ensemble, and it is proposed that the steady states in integrable systems are described by the generalized Gibbs ensemble~\cite{PhysRevLett.98.050405,Essler_2016}, whose density matrix is constructed from an extensive number of local and quasi-local conserved quantities as well as the Hamiltonian~\cite{PhysRevLett.115.120601,PhysRevLett.115.157201}. The second example is the generalized hydrodynamics~\cite{PhysRevX.6.041065,PhysRevLett.117.207201}, which describes large scale nonequilibrium dynamics in integrable systems and is formulated from the set of continuity equations for conserved quantities. 
In many interacting integrable systems which are solved by the Bethe ansatz and the quantum inverse scattering methods~\cite{korepin_bogoliubov_izergin_1993,baxter2007exactly}, the existence of local conserved quantities and the mutual commutativity of them were proved from the commutativity of transfer matrices $T\paren{\lambda}$ with different values of the spectral parameter $\lambda$: $[T\paren{\lambda},T\paren{\mu}]=0$. Local conserved quantities are obtained from the expansion of $\ln T\paren{\lambda}$ in terms of $\lambda$, which includes the Hamiltonian. Another standard method to construct local conserved quantities is to use the boost operator $B$~\cite{tetelman1982lorentz,10.1143/PTP.69.431,THACKER1986348}. In this method, local conserved quantities are obtained recursively from the commutation relations as $[B, Q_{n}]=Q_{n+1}$.

Although how to prove the existence of local conserved quantities and construct them are known, it is still difficult to obtain the explicit expression for them because the calculation is complicated in general, and one needs to find the pattern of coefficients of local conserved quantities to express general local conserved quantities. Grabowski and Mathieu investigated the problem for the XYZ spin-$1/2$ chain, which is a generalization of the Heisenberg spin-$1/2$ chain and known as an integrable spin 
chain~\cite{baxter2007exactly,mccoy1968hydrogen,sutherland1970two,PhysRevLett.26.832,PhysRevLett.26.834,BAXTER1972193,BAXTER1972323,BAXTER19731,BAXTER197325,BAXTER197348} with the use of the boost operator. As a result, they found the explicit expression in the case of the Heisenberg chain (also called the XXX chain)~\cite{grabowski1994quantum,GRABOWSKI1995299}. In the more general case, so far, $Q_{n}$ has been obtained only in the case of $3\leq n \leq 6$ from the Hamiltonian $Q_{2}=H$, and the explicit expression for general local conserved quantities was not found.
For the study of nonequilibrium phenomena, the explicit expression for local conserved quantities is a useful tool.
For example, current operators, which are fundamental ones in the study of transport phenomena~\cite{PhysRevB.55.11029}, can be constructed from the continuity equations of the densities of local conserved quantities. 

In this Letter, we present an explicit expression for all the local conserved quantities in the XYZ spin-$1/2$ chain with general coupling constants. 
We also confirm that our result can be applied to the XXZ spin-$1/2$ chain with a magnetic field in the z-axis direction, which is also known as a Bethe ansatz soluble model~\cite{korepin_bogoliubov_izergin_1993,takahashi_1999}.
To obtain the expression, we have used a straightforward way with a notation called \textit{doubling-product}, which was introduced to prove the absence of local conserved quantities in the XYZ spin-$1/2$ chain with a magnetic field~\cite{shiraishi2019proof} and its extension. 
We have directly derived the conditions for the commutator of each local conserved quantity and the Hamiltonian to be zero.
With the doubling-product notation, we have found the pattern of coefficients of local conserved quantities and obtained general solutions of them.
 
\textit{Model and $k$-support conserved quantities.---}
We consider the XYZ spin-$1/2$ chain without a magnetic field for periodic boundary conditions:
\begin{align}
H=\sum_{i=1}^{L}\paren{J_{X}X_{i}X_{i+1}+J_{Y}Y_{i}Y_{i+1}+J_{Z}Z_{i}Z_{i+1}},
\end{align}
where $X_{i}$, $Y_{i}$, and $Z_{i}$ represent the Pauli matrices $\sigma^{x}$, $\sigma^{y}$, and $\sigma^{z}$ acting on the spin at the site $i$, respectively. We set all the coupling constants $J_{X}$, $J_{Y}$, and $J_{Z}$ nonzero.  Following Ref.~\cite{shiraishi2019proof}, we define $k$-support conserved quantities $Q_{k}$:
\begin{align}
Q_{k}=\sum_{l=1}^{k}\sum_{\bm{A}^{l}}\sum_{i=1}^{L}q_{\bm{A}^{l}}\bm{A}_{i}^{l}.
\end{align}
Here, $\bm{A}_{i}^{l}\equiv A_{i}^{1}A_{i+1}^{2}\cdots A_{i+l-1}^{l}$ is a sequence of $l$ operators acting from the site $i$ to the site $i+l-1$. Operators at both ends $A^{1}$, $A^{l}$ take $X$, $Y$, or $Z$, and the other operators $A^{2},\ldots, A^{l-1}$ take $X$, $Y$, $Z$, or the identity operator $I$. $\sum_{i}\bm{A}_{i}^{l}$ is called an $l$-support operator. Coefficients $\bce{q_{\bm{A}^{l}}}$ are determined from the commutation relation $\left[Q_{k},H\right]=0$.
For example, the Hamiltonian itself is a trivial $2$-support conserved quantity, and it is easily proved that all the $1$-support conserved quantities 
are $\sum_{i}X_{i}$ if $J_{Y}=J_{Z}$, $\sum_{i}Y_{i}$ if $J_{Z}=J_{X}$, and $\sum_{i}Z_{i}$ if $J_{X}=J_{Y}$. Therefore, we consider $Q_{k}$ for $k\geq 2$ hereafter, and our aim is to determine the coefficients $\bce{q_{\bm{A}^{l}}}$ of $Q_{k}$.  

To describe commutation relations, we use the following notation~\cite{shiraishi2019proof}:
\begin{align}
&\begin{array}{cccc}
X_{i}&Y_{i+1}&Z_{i+2}&\\
&&X_{i+2}&X_{i+3}
\end{array}\nonumber\\
\equiv&-\imi\left[X_{i}Y_{i+1}Z_{i+2},X_{i+2}X_{i+3}\right]/2\nonumber\\
=&X_{i}Y_{i+1}Y_{i+2}X_{i+3},
\end{align}
and we drop the subscripts hereafter for visibility. Fundamental formulae using this notation are
\begin{align}
&\begin{array}{rcc}
&X&Y\\
&X&X\\
=-&I&Z,
\end{array}
\quad
\begin{array}{rcc}
&X&Y\\
&Y&Y\\
=&Z&I,
\end{array}
\quad
\begin{array}{rcc}
&X&Y\\
&Z&Z\\
=&0,&
\end{array}\label{eq:com1}\\
&\ \ \ \begin{array}{rcc}
&X&X\\
&X&X\\
=&0,&
\end{array}
\quad
\begin{array}{rcc}
&X&X\\
&Y&Y\\
=&0,&
\end{array}
\quad
\begin{array}{rcc}
&X&X\\
&Z&Z\\
=&0,&
\end{array}\label{eq:com2}\\
&\ \ \ \begin{array}{rcc}
&X&I\\
&X&X\\
=&0,&
\end{array}
\quad
\begin{array}{rcc}
&X&I\\
&Y&Y\\
=&Z&Y,
\end{array}
\quad
\begin{array}{rcc}
&X&I\\
&Z&Z\\
=-&Y&Z.\label{eq:com3}
\end{array}
\end{align}

\textit{Doubling-product operators and their extension---}
First we consider the case that the site number $L$ satisfies $k\leq L/2$. As shown in Ref.~\cite{shiraishi2019proof}, by considering $(k+1)$-support operators in $\left[Q_{k},H\right]$, $k$-support operators in $Q_{k}$ are restricted to doubling-product operators defined as
\begin{align}
&~~~~~\overline{A_{1}A_{2}\cdots A_{k-2}A_{k-1}}\nonumber\\
&\begin{array}{ccccccc}
=~c&A_{1}&\paren{A_{1}A_{2}}&\paren{A_{2}A_{3}}&\cdots&\paren{A_{k-2}A_{k-1}}&A_{k-1}\\
=~~&A_{1}&A_{1,2}&A_{2,3}&\cdots &A_{k-2,k-1}&A_{k-1},
\end{array}
\label{eq:doubling}
\end{align}
where $A_{\alpha}$ takes one of $\bce{X,Y,Z}$ and it is required that $A_{\alpha}\neq A_{\alpha+1}$. We define $A_{\alpha,\beta}$ by $\bce{A_{\alpha},A_{\beta},A_{\alpha,\beta}}=\bce{X,Y,Z}$ when $A_{\alpha}\neq A_{\beta}$. The coefficient $c\in\bce{\pm1,\pm\imi}$ is determined from Eq.~\eqref{eq:doubling}.
Furthermore, after fixing a normalization factor of $Q_{k}$, nonzero coefficients of $k$-support operators are uniquely given by
\begin{align}
&q_{\overline{A_{1}A_{2}\cdots A_{k-2}A_{k-1}}}\nonumber\\
=~&s\paren{A_{1}A_{2}\cdots A_{k-2}A_{k-1}}J_{A_{1}}J_{A_{2}}\cdots J_{A_{k-2}}J_{A_{k-1}},
\label{eq:k-0-c}
\end{align}
where $s\paren{XY}=s\paren{YZ}=s\paren{ZX}=-s\paren{YX}=-s\paren{ZY}=-s\paren{XZ}\equiv1$, and $s\paren{A_{1}A_{2}\cdots A_{k-2}A_{k-1}}\equiv s\paren{A_{1}A_{2}}s\paren{A_{2}A_{3}}\cdots s\paren{A_{k-2}A_{k-1}}$.
Therefore, for $2\leq k\leq L/2$, $Q_{k}$ is unique up to differences of smaller support conserved quantities $Q_{k^{\prime}<k}$. Note that $Q_{k}+Q_{k^{\prime}<k}$ is also a $k$-support conserved quantity.

To express $k^{\prime}(<k)$-support operators in $Q_{k}$, it is useful to extend the definition of doubling-product operators.
Let us allow the case that neighboring symbols in doubling-product operators are the same $A_{\alpha}= A_{\alpha+1}$. Then, in the definition Eq.~\eqref{eq:doubling}, $A_{\alpha,\alpha +1}$ is replaced by $I$ if $A_{\alpha}= A_{\alpha+1}$. When the condition $A_{\alpha}= A_{\alpha+1}$ satisfies at $m$ places  in an $l$-support operators, we call it an $(l, m)$ operator. $m$ is called the number of \textit{holes} and used to study the structure of conserved quantities~\cite{grabowski1994quantum,GRABOWSKI1995299}. Under this definition, all the $k$-support operators in $Q_{k}$ are $(k,0)$ operators.

We can express $(l, m)$ operators as 
\begin{align}
&\overline{\underbrace{A_{1}\cdots A_{1}}_{1+m_{1}}\underbrace{A_{2}\cdots A_{2}}_{1+m_{2}}\cdots\cdots \underbrace{A_{l-m-1}\cdots A_{l-m-1}}_{1+m_{l-m-1}}}\nonumber\\
\equiv~& \overline{A_{1}^{1+m_{1}}A_{2}^{1+m_{2}}\cdots A_{l-m-1}^{1+m_{l-m-1}}},
\label{eq:lmops}
\end{align}
where $A_{\alpha}\neq A_{\alpha+1}$ and $m_{j}\geq 0$ is an integer which satisfies $\sum_{j=1}^{l-m-1}m_{j}=m$.
For example, $\overline{X^{2}Z^{2}}=\overline{XXZZ}=XIYIZ$ and $\overline{X^{3}Z}=\overline{XXXZ}=XIIYZ$ are both $(5,2)$ operators. 
When we consider commutation relations of $(l, m)$ operators, we use the following notation
\begin{align}
&\begin{array}{cccc}
\cline{1-3}
X&Y&Z^{2}\\
\cline{2-2}
&Z&&
\end{array}
\equiv
\begin{array}{ccccc}
X&Z&X&I&Z\\
&Z&Z&&
\end{array}\nonumber\\
=&
\begin{array}{rccccc}
-&X&I&Y&I&Z
\end{array}
=
\begin{array}{rcccc}
\cline{2-3}
-&X^{2}&Z^{2},&&
\end{array}
\end{align}
where $\overline{XYZ^{2}}$ is a $(5,1)$ operator, and  $\overline{X^{2}Z^{2}}$ is a $(5,2)$ operator.


\begin{figure}[tb]
\centering
\includegraphics[width=0.8\columnwidth,clip]{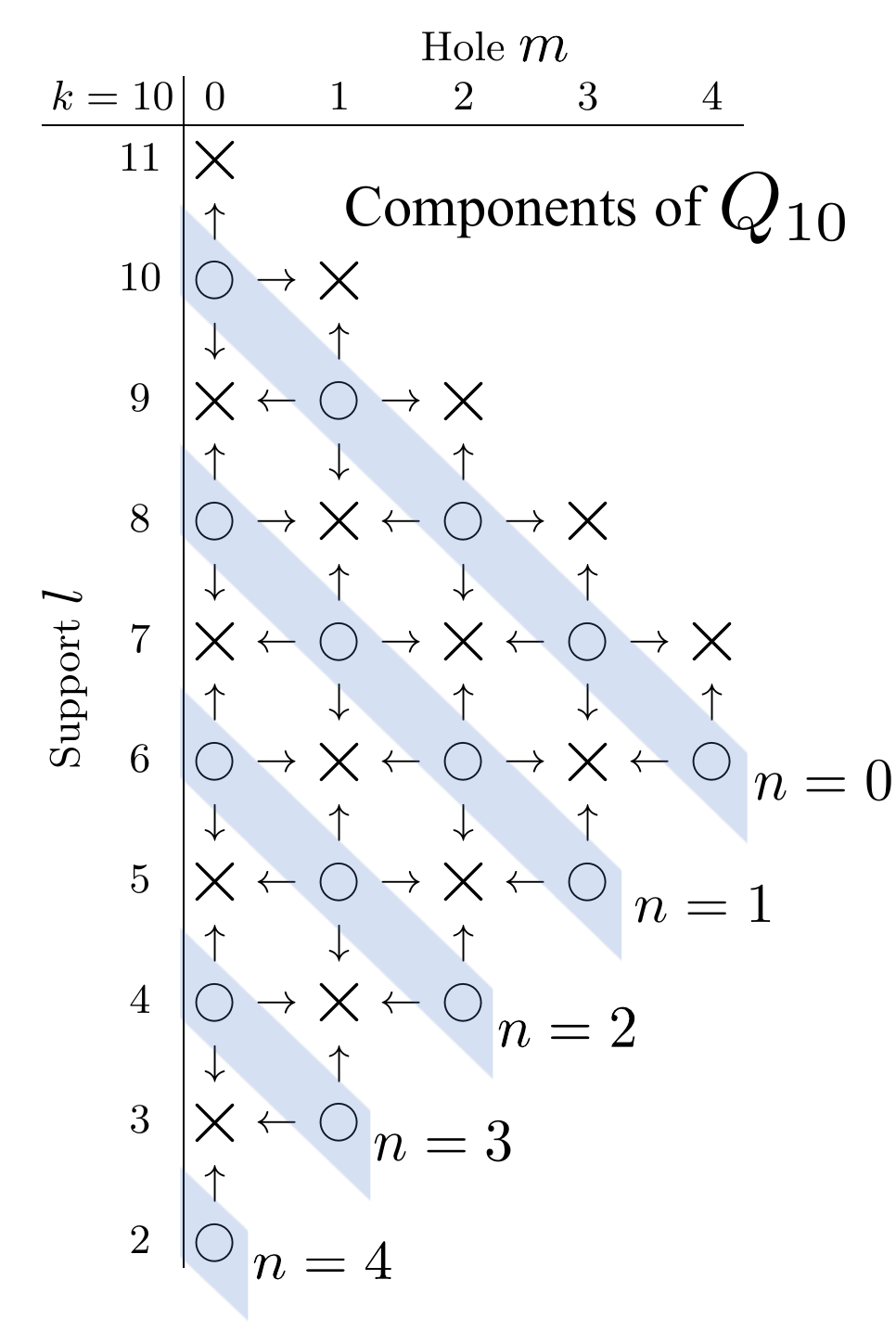}
\caption{Structure of a $k$-support conserved quantity $Q_{k}$ for $k=10$. Circles represent $(l, m)$ operators in $Q_{k}$, where $l=k-2n-m$. Crosses represent operators generated by the commutation relation of $H$ and operators represented as circles, which are to be cancelled.}
\label{fig:structure}
\end{figure}

\textit{Structure of $Q_{k}$.---}
Let us consider the commutation relation of an $(l, m)$ operator in $Q_{k}$ $\overline{A_{1}^{1+m_{1}}A_{2}^{1+m_{2}}\cdots A_{l-m-1}^{1+m_{l-m-1}}}$ and $H$.
Candidates of operators in the commutator are $(l\pm 1)$ and $l$-support. First, $(l-1)$-support operators are constructed by removing $A_{1}$ or $A_{l-m-1}$. As for $A_{1}$, this operator is
\begin{align}
\begin{array}{ccccc}
\cline{1-5}
A_{1}&A_{1}^{m_{1}}&A_{2}^{1+m_{2}}&\cdots&A_{l-m-1}^{1+m_{l-m-1}}\\
\cline{1-1}
A_{1}
\end{array}.
\end{align}
Note that this term is nonzero only if $m_{1}=0$ and $A_{1}\neq A_{2}$. Therefore, the number of holes is conserved, and it is an $(l-1,m)$ operator. The same holds for $A_{l-m-1}$. Second, $(l+1)$-support operators are constructed by adding $A_{0}\paren{\neq A_{1}}$ on the left side of $A_{1}$:
\begin{align}
\begin{array}{ccccc}
\cline{2-5}
&A_{1}^{1+m_{1}}&A_{2}^{1+m_{2}}&\cdots&A_{l-m-1}^{1+m_{l-m-1}}\\
\cline{1-1}
A_{0}&&&&
\end{array},
\end{align}
or $A_{l-m}\paren{\neq A_{l-m-1}}$ on the right side of $A_{l-m-1}$. Therefore, these operators are $(l+1,m)$ operators.
The third case of $l$-support operators is a bit more complicated. 
For example, they are given as
\begin{align}
\begin{array}{ccccc}
\cline{1-5}
A_{1}^{1+m_{1}}&\cdots&A_{p}&\cdots&A_{l-m-1}^{1+m_{l-m-1}}\\
\cline{3-3}
&&B_{p}&&
\end{array}.
\end{align}
If $A_{p}=B_{p}$ for $1<p<l-1$, these operators cannot be expressed as $(l, m)$ operators. However, these terms are cancelled and do not contribute to the commutator. In the case of $A_{p}\neq B_{p}$, from Eqs.~\eqref{eq:com1}-\eqref{eq:com3}, only $(l, m\pm1)$ operators are obtained (see Supplemental Material~\cite{foot:supplemental} for the details).

Consequently, operators in $Q_{k}$ are classified as $(l, m)$ operators as shown in Fig.~\ref{fig:structure}. Here, we fix the degrees of freedom to add $Q_{k^{\prime}<k}$. For example, coefficients of $(k-2n-1,0)$ operators ($n=0,1,\ldots$) are set to zero. In Fig.~\ref{fig:structure}, circles represent $(l, m)$ operators in $Q_{k}$, and crosses shown by arrows represent operators generated by the commutation relations of $(l, m)$ operators in $Q_{k}$ and $H$. 
From this structure, $Q_{k}$ is represented as
\begin{align}
Q_{k}=\sum_{\substack{0\leq n+m \leq \lfloor \frac{k}{2}\rfloor-1,\\ n,m\geq 0}}\sum_{\substack{\overline{\bm{A}}:\\
(k-2n-m,m)~\text{operators}}}q^{k-2n-m,m}_{\overline{\bm{A}}}\ \overline{\bm{A}}.
\label{eq:Qqlm}
\end{align}
Here, the sum of $\overline{\bm{A}}\equiv\overline{A_{1}^{1+m_{1}}A_{2}^{1+m_{2}}\cdots A_{k-2n-2m-1}^{1+m_{k-2n-2m-1}}}$ runs over all $(k-2n-m, m)$ operators that satisfy $n\geq 0$, $m\geq 0$, and  $k-2n-2m\geq 2$, which corresponds to circles in Fig.~\ref{fig:structure}. 

\textit{Recursive way.---}
One of the main results of this Letter is that we have found a simple recursive way to determine all the coefficients $\{q^{k-2n-m,m}_{\overline{\bm{A}}}\}$ in Eq.~\eqref{eq:Qqlm}~\cite{foot:supplemental}. By introducing some functions, we describe the way.

First, $q^{k-2n-m,m}_{\overline{\bm{A}}}$ is expressed using the function $R$ as
\begin{align}
&q^{w,m}_{\overline{A_{1}^{1+m_{1}}A_{2}^{1+m_{2}}\cdots A_{t}^{1+m_{t}}}}
=s\paren{A_{1}A_{2}\cdots A_{t}}\paren{J_{X}J_{Y}J_{Z}}^{m}
\nonumber\\
&\times \paren{\prod_{j=1}^{t}J_{A_{j}}^{1-m_{j}}} R^{w, m}\paren{A_{1}A_{2}\cdots A_{t}},
\label{eq:qandr}
\end{align}
where we introduce the notation $w\equiv k-2n-m$, $t\equiv k-2n-2m-1$ for simplicity. $A_{1}A_{2}\cdots A_{t}$ is a character string of length $t\geq 1$, and $A_{1}$, $A_{2}$, \ldots, $A_{t}$ take one of $\bce{X,Y,Z}$, respectively. 
$s$ is the function we introduced in Eq.~\eqref{eq:k-0-c}. 
For example, $q^{5,2}_{\overline{X^{2}Z^{2}}}=(J_{X}J_{Y}J_{Z})^{2}R^{5,2}(XZ)$ and $q^{5,2}_{\overline{X^{3}Z}}=(J_{X}J_{Y}J_{Z})^{2}(J_{Z}/J_{X})R^{5,2}(XZ)$.
We remark that $R$ does not depend on where holes are because it does not depend on $m_{1},m_{2},\ldots,m_{t}$. 

Second, $R$ is represented as a linear combination of the function $S$: 
\begin{align}
R^{w,m}(A_{1}A_{2}\cdots A_{t})=\sum_{\tilde{n}=0}^{n}g^{w,m}_{n-\tilde{n}}S_{\tilde{n}}(A_{1}A_{2}\cdots A_{t}), 
\label{eq:rands}
\end{align}
where $S$ is defined as $S_{0}\paren{A_{1}A_{2}\cdots A_{t}}\equiv 1$ and 
\begin{align}
S_{\tilde{n}}\paren{A_{1}A_{2}\cdots A_{t}}\equiv \sum_{1\leq j1\leq j2\leq\cdots\leq j\tilde{n}\leq t} J^{2}_{A_{j1}}J^{2}_{A_{j2}}\cdots J^{2}_{A_{j\tilde{n}}}
\label{eq:sp}
\end{align}
for $\tilde{n}\geq 1$, and $g^{w,m}_{n-\tilde{n}}~(0\leq \tilde{n}\leq n)$ does not depend on $A_{1}A_{2}\cdots A_{t}$. 
By definition, $R^{w,m}$ is a symmetric polynomial in $J^{2}_{A_{1}},J^{2}_{A_{2}},\ldots , J^{2}_{A_{t}}$.

\begin{figure}[tb]
\centering
\includegraphics[width=\columnwidth,clip]{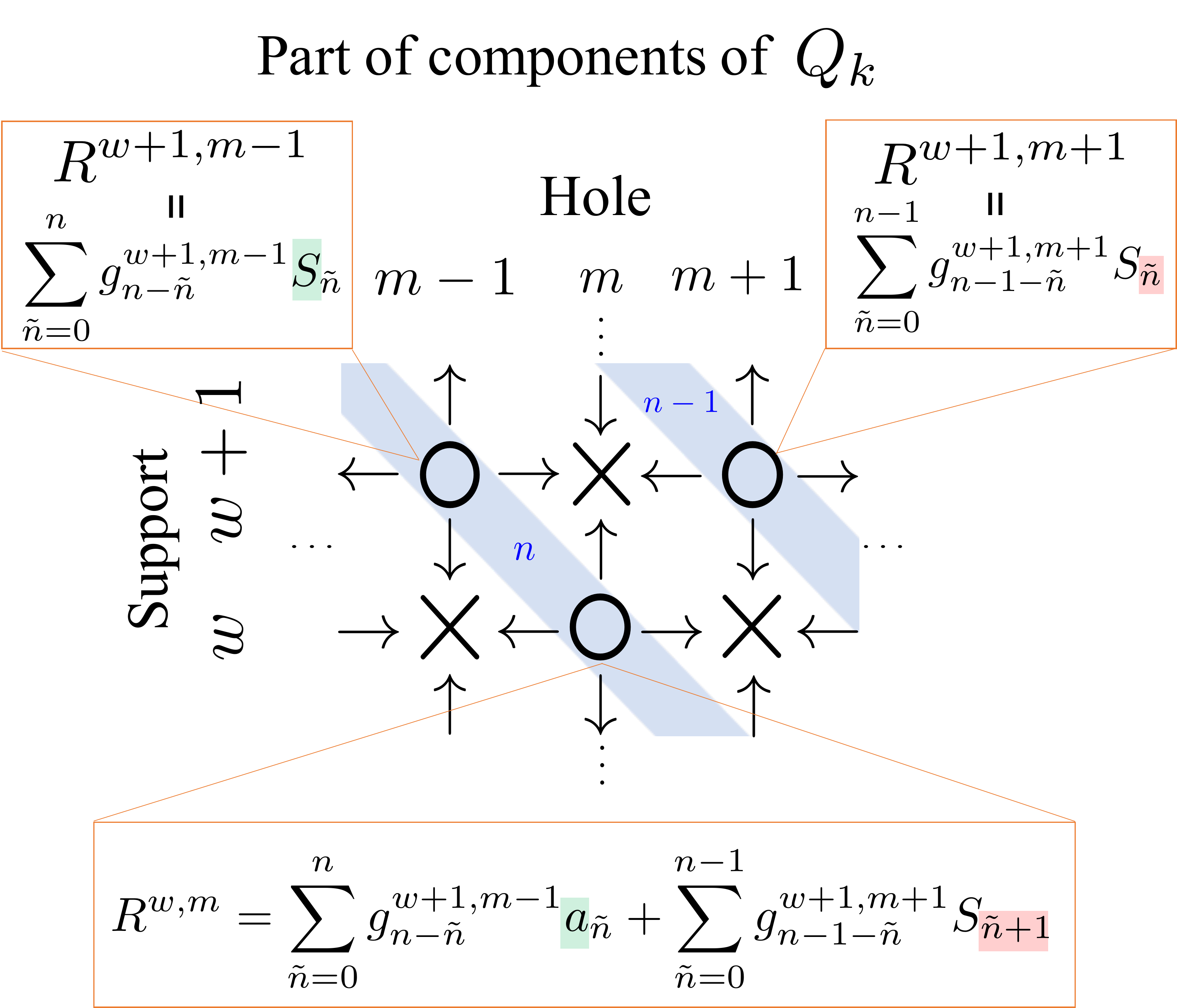}
\caption{Recursive way to obtain the function $R^{w, m}$ in Eq.~\eqref{eq:qandr}. When $R^{w+1, m-1}=\sum_{\tilde{n}=0}^{n}g^{w+1,m-1}_{n-\tilde{n}}S_{\tilde{n}}$ and $R^{w+1, m+1}=\sum_{\tilde{n}=0}^{n-1}g^{w+1,m+1}_{n-1-\tilde{n}}S_{\tilde{n}}$ are determined, then $R^{w, m}$ is determined as the sum of two terms; the first term is obtained by the replacement $S_{\tilde{n}}$ to $a_{\tilde{n}}$ in $R^{w+1, m-1}$, and the second term is obtained by the replacement $S_{\tilde{n}}$ to $S_{\tilde{n}+1}$ in $R^{w+1, m+1}$.}
\label{fig:procedure}
\end{figure}

Finally, $g^{w,m}_{n-\tilde{n}}$ is determined as follows. From Eq.~\eqref{eq:k-0-c}, which corresponds to the case of $n=0$, $g^{k-m,m}_{0}=1$ because $R^{k-m,m}=1$. Figure~\ref{fig:procedure} shows how to determine the other $g^{w,m}_{n-\tilde{n}}$'s recursively.
Suppose that $R^{w+1,m-1}=\sum_{\tilde{n}=0}^{n}g^{w+1,m-1}_{n-\tilde{n}}S_{\tilde{n}}$ (the upper left circle) and $R^{w+1,m+1}=\sum_{\tilde{n}=0}^{n-1}g^{w+1,m+1}_{n-1-\tilde{n}}S_{\tilde{n}}$ (the upper right circle) are determined.
Then $R^{w, m}$ (the lower center circle) is determined as $\sum_{\tilde{n}=0}^{n}g^{w+1,m-1}_{n-\tilde{n}}a_{\tilde{n}}+\sum_{\tilde{n}=0}^{n-1}g^{w+1,m+1}_{n-1-\tilde{n}}S_{\tilde{n}+1}$. 
Here, in the case of $m=0$, the term with respect to $R^{w+1,m-1}$ is regarded as zero. $a_{n}$ is defined as 
\begin{widetext}
\begin{gather}
a_{n}\equiv \frac{J^{2}_{X}(J^{2(n+2)}_{Y}-J^{2(n+2)}_{Z})+J^{2}_{Y}(J^{2(n+2)}_{Z}-J^{2(n+2)}_{X})+J^{2}_{Z}(J^{2(n+2)}_{X}-J^{2(n+2)}_{Y})}{(J^{2}_{X}-J^{2}_{Y})(J^{2}_{Y}-J^{2}_{Z})(J^{2}_{Z}-J^{2}_{X})}.
\label{eq:defa}
\end{gather}
\end{widetext}
By following this recursive way, all the other $g^{w,m}_{n-\tilde{n}}$'s are determined. 
$a_{n}$ is characterized as the coefficient of $u^{2}$ in the remainder
of the division of a monomial $u^{n+2}$ by $(u-J^{2}_{X})(u-J^{2}_{Y})(u-J^{2}_{Z})$, and plays an important role in our proof that this recursive way is correct~\cite{foot:supplemental}. We note that even if $J_{X}=J_{Y}$, $a_{n}$ does not diverge by the characterization of $a_{n}$. $a_{n}$ is also obtained from the recurrence relation $a_{n+3}=\paren{J^{2}_{X}+J^{2}_{Y}+J^{2}_{Z}}a_{n+2}-\paren{J^{2}_{X}J^{2}_{Y}+J^{2}_{Y}J^{2}_{X}+J^{2}_{Z}J^{2}_{
X}}a_{n+1}+J^{2}_{X}J^{2}_{Y}J^{2}_{Z}a_{n}$, $a_{-2}=a_{-1}=0$, and $a_{0}=1$. 

In the case of the XXX chain ($J_{X}=J_{Y}=J_{Z}=1$), the pattern of $R^{w, m}$ becomes more simple~\cite{foot:supplemental}. It is satisfied that $R^{w, m}=R^{w+1,m-1}+R^{w+1,m+1}$ for $m\geq 1$, which reproduces the known structure called a Catalan tree in 
Refs.~\cite{grabowski1994quantum,GRABOWSKI1995299}.

\textit{Explicit expression for $Q_{k}$.---}
By solving the recursive way discussed above explicitly, we have obtained the explicit expression for $Q_{k}$~\cite{foot:supplemental}. The solution for $g^{w,m}_{n-\tilde{n}}\equiv g^{k-2n-m,m}_{n-\tilde{n}}$ is $k$-independent and given by
\begin{align}
g^{w,m}_{n-\tilde{n}}
=f\paren{n-\tilde{n},m+\tilde{n}},
\label{eq:defr}
\end{align}
where $f$ is defined as $f\paren{0,m}\equiv 1$ and
\begin{align}
&f\paren{n,m}\nonumber\\
\equiv& \frac{m}{n+m}\sum_{p=1}^{n}\binom{n+m}{p}\sum_{\substack{j1,j2,\ldots,jp\geq 1\\\\j1+j2+\cdots+jp=n}}a_{j1}a_{j2}\cdots a_{jp}
\label{eq:deff}
\end{align}
for $n\geq 1$.

For $k\leq 6$, the explicit expression for $Q_{k}$ was calculated in Ref.~\cite{GRABOWSKI1995299}. Here, as an example, we present the coefficients of $0$-hole operators in $Q_{8}$. They are given as $q^{8,0}\paren{\overline{A_{1}A_{2}\cdots A_{7}}}=s\paren{A_{1}A_{2}\cdots A_{7}}\prod_{j=1}^{7}J_{A_{j}}$, $q^{6,0}\paren{\overline{A_{1}A_{2}\cdots A_{5}}}=s\paren{A_{1}A_{2}\cdots A_{5}}\paren{\prod_{j=1}^{5}J_{A_{j}}}\sum_{j=1}^{5}J_{A_{j}}^{2}$, $q^{4,0}\paren{\overline{A_{1}A_{2}A_{3}}}=s\paren{A_{1}A_{2}A_{3}}\paren{\prod_{j=1}^{3}J_{A_{j}}} [J_{A_{1}}^{4}+J_{A_{2}}^{4}+J_{A_{3}}^{4}+J_{A_{1}}^{2}J_{A_{2}}^{2}+J_{A_{2}}^{2}J_{A_{3}}^{2}+J_{A_{3}}^{2}J_{A_{1}}^{2}+\paren{J_{X}^{2}+J_{Y}^{2}+J_{Z}^{2}}\sum_{j=1}^{3}J^{2}_{A_{j}}]$, and 
$q^{2,0}\paren{\overline{A_{1}}}=J_{A_{1}}[J_{A_{1}}^{6}+2\paren{J^{2}_{X}+J^{2}_{Y}+J^{2}_{Z}}J_{A_{1}}^{4}+
\paren{2J^{4}_{X}+2J^{4}_{Y}+2J^{4}_{Z}+3J^{2}_{X}J^{2}_{Y}+3J^{2}_{Y}J^{2}_{Z}+3J^{2}_{Z}J^{2}_{X}}J^{2}_{A_{1}}]$.

Let us compare some properties of $Q_{2k^{\prime}}$ and $Q_{2k^{\prime}+1}$, where $k^{\prime}$ is a positive integer. We first consider the time-reversal symmetry.
$Q_{2k^{\prime}(+1)}$ consists of the sum of $(2k^{\prime}(+1)-2n-m, m)$ operators as shown in Eq.~\eqref{eq:Qqlm}. Since $(2k^{\prime}(+1)-2n-m, m)$ operators act on an even (odd) number of sites as the Pauli matrices, these operators are (anti-)symmetric under time-reversal. Therefore, $Q_{2k^{\prime}(+1)}$ is (anti-)symmetric under time-reversal.
We next consider the similarity between $Q_{2k^{\prime}}$ and $Q_{2k^{\prime}+1}$.
Since $\lfloor \frac{2k^{\prime}}{2}\rfloor=\lfloor \frac{2k^{\prime}+1}{2}\rfloor=k^{\prime}$, $n$ and $m$ included in the summation of $Q_{2k^{\prime}}$ and $Q_{2k^{\prime}+1}$ in Eq.~\eqref{eq:Qqlm} take the same values. For this reason, the coefficients of the operators in $Q_{2k^{\prime}}$ and $Q_{2k^{\prime}+1}$ have a similar structure. 

We note that even if one or two coupling constants are zero, $Q_{k}$ we obtained can be used. However, $Q_{k}$ may be a conserved quantity multiplied by coupling constants set to zero, and it needs to be divided by the coupling constants in this case. Another difference 
from the case that all the coupling constants are nonzero is that $k$-support conserved quantities are not unique even for $2\leq k\leq L/2$.
In fact, it is known that there is another family of local conserved quantities in addition to $Q_{k}$'s~\cite{grady_1982,gusev_1982,itoyama_1989,araki_1990,GRABOWSKI1995299}. For example, in the case of $J_{Z}=0$, $\sum_{i}(X_{i}Y_{i+1}-Y_{i}X_{i+1})$ is another $2$-support conserved quantity in addition to the Hamiltonian $Q_{2}=H$ itself. 

\textit{Commutativity with a magnetic field in the case of the XXZ chain.---}
In the case of $J_{X}=J_{Y}$, we confirm that $[Q_{k}, \sum_{i}Z_{i}]=0$, \textit{i.e.}, $Q_{k}$ is also conserved in the XXZ spin-$1/2$ chain with a magnetic field in the z-axis direction. Since it is known that the transfer matrix of the XXZ spin-$1/2$ chain commutes with the magnetic field $[T(\lambda),\sum_{i}Z_{i}]=0$~\cite{korepin_bogoliubov_izergin_1993}, local conserved quantities obtained from the expansion of $\ln T(\lambda)$ in terms of $\lambda$ also commute with the magnetic field. In addition, the uniqueness of $k$-support conserved quantities for $2\leq k\leq L/2$ also holds in the presence of the magnetic field because commutation relations of the magnetic field and $k$-support operators generate no $(k+1)$-support operators. Therefore, $Q_{k}$ is a linear combination of local conserved quantities obtained from the transfer matrix method, and $[Q_{k},\sum_{i}Z_{i}]=0$ is satisfied. We note that one can prove this commutativity explicitly by using Eqs.~\eqref{eq:Qqlm}-\eqref{eq:deff}~\cite{foot:supplemental}.

\textit{Case of $L/2<k\leq L$.---}
In the case of $L/2<k\leq L$, a different point from the case of $2\leq k\leq L/2$ is that commutators of different support operators can be cancelled. In this case, the conditions we impose in the above discussion for $2\leq k\leq L/2$ become not necessary but sufficient for $[Q_{k}, H]=0$. Therefore, $Q_{k}$ we obtain is also conserved for $L/2<k\leq L$, although it is not necessarily the unique $k$-support conserved quantity.

\textit{Summary and Outlook.---}
We have presented the rigorous explicit expression for $k$-support conserved quantities in the XYZ spin-$1/2$ chain $\bce{Q_{k}}$ for $1\leq k\leq L$. Doubling product is a useful notation to find and express them. By using this notation, we have derived a recursive way to obtain the coefficients of $Q_{k}$ directly and have found the solution. We have also confirmed that $Q_{k}$  for $k\geq 2$ is conserved even in the case of the XXZ model with a magnetic field in the z-axis direction. 
The XXX chain was the exceptional case that the expression is known~\cite{grabowski1994quantum,GRABOWSKI1995299}, and our result has expanded the scope to general coupling constants. In particular, it enables us to analyze the coupling constants dependence of the local conserved quantities. Once the expression Eqs.~\eqref{eq:Qqlm}-\eqref{eq:deff} is obtained, one can easily handle the local conserved quantities both analytically and numerically. Therefore, our result may serve as a new tool for the study of nonequilibrium phenomena in interacting integrable spin chains.

\textit{Acknowledgements.---}
We thank H.~Tsunetsugu and H.~Katsura for fruitful discussion and comments on the manuscript.
K.~F.~acknowledges support by the Forefront Physics and Mathematics Program to Drive Transformation (FoPM), WINGS Program, the University of Tokyo.

\bibliographystyle{unsrt}



\clearpage
\setcounter{equation}{0}
\makeatletter
\def\tagform@#1{\maketag@@@{(S#1)}}
\makeatother
\setcounter{figure}{0}
\renewcommand{\thefigure}{S\arabic{figure}}

\makeatletter
\renewcommand{\@cite}[2]{[S#1]}
\renewcommand{\@biblabel}[1]{[S#1]}
\makeatother

\onecolumngrid
\vspace{0.5cm}
\begin{center} 
{\Large {\bf Supplemental Material  for ``Explicit Construction of Local Conserved Quantities in the XYZ Spin-1/2 Chain''}} 
\\
\vspace{0.3cm}
Yuji Nozawa$ ^{\ast}$ and Kouhei Fukai\\
The Institute for Solid State Physics, The University of Tokyo, Kashiwa, Chiba 277-8581, Japan

\end{center}

\section*{S1. Derivation of $k$-support conserved quantities}
We derive $k$-support conserved quantities $Q_{k}$ for $k\geq 3$. Here, we assume that $k\leq L/2$, although $Q_{k}$ obtained under the assumption is conserved for the case of $L/2<k\leq L $ as discussed in the main text. We prove that all operators in $Q_{k}$ are $(l,m)$ operators defined in the main text and derive that all coefficients of the operators are expressed as 
Eqs.~(15)-(20).
Our proof is organized as follows. First, we derive coefficients of $(k-m, m)$ operators for $m>1$ by using that of $(k, 0)$ operators
Eq.~(8).
Then we derive conditions that the other coefficients satisfy for $Q_{k}$ to be conserved, and we present a recursive way to construct these coefficients. Finally, we derive them in closed form as shown in 
Eqs.~(14)-(20).
For later use, we define the function $r$ as
\begin{align}
q^{k-2n-m,m}_{\overline{A_{1}^{1+m_{1}}A_{2}^{1+m_{2}}\cdots A_{k-2n-2m-1}^{1+m_{k-2n-2m-1}}}}
\equiv s\paren{A_{1}A_{2}\cdots A_{k-2n-2m-1}}\paren{J_{X}J_{Y}J_{Z}}^{m}\paren{\prod_{j=1}^{k-2n-2m-1}J_{A_{j}}^{1-m_{j}}}
\nonumber\\
\times r^{k-2n-m, m}\paren{A_{1}^{1+m_{1}}A_{2}^{1+m_{2}}\cdots A_{k-2n-2m-1}^{1+m_{k-2n-2m-1}}}.
\label{supeq:defr}
\end{align}
In this section, we prove that 
\begin{align}
r^{k-2n-m, m}\paren{A_{1}^{1+m_{1}}A_{2}^{1+m_{2}}\cdots A_{k-2n-2m-1}^{1+m_{k-2n-2m-1}}}=R^{k-2n-m, m}\paren{A_{1}A_{2}\cdots A_{k-2n-2m-1}}, 
\end{align}
where the function $R$ is given in 
Eqs.~(16)-(20).

For later convenience, we define
\begin{align}
\wsup\equiv k-2n-m,\quad \knm\equiv k-2n-2m-1
\label{supeq:knmdef}
\end{align}
as in the main text.

\subsection{A. Coefficients of $(k-m, m)$ operators}
In this subsection, we calculate the coefficients of $(k-m, m)$ operators in $Q_{k}$. As shown in Ref.~\cite{shiraishi2019proof}, all the $k$-support operators in $Q_{k}$ are $(k,0)$ operators.
We show commutators of $(k,0)$ operators and $H$ generate only $(k-1,0)$ and $(k,1)$ operators.

Let $\overline{A_{1}A_{2}\cdots A_{k-1}}$ be an arbitrary $(k,0)$ operator. By definition, $A_{p}\neq A_{p+1}$ for all $1\leq p \leq k-2$.
From the commutator of $\overline{A_{1}A_{2}\cdots A_{k-1}}$ and $H$, $(k-1)$-support operators are only generated by
\begin{align}
\begin{array}{cccc}
\cline{1-4}
A_{1}&A_{2}&\cdots&A_{k-1}\\
\cline{1-1}
A_{1}
\end{array}=
\begin{array}{rcccc}
\cline{2-4}
s\paren{A_{1,2}A_{1}}&A_{2}&\cdots&A_{k-1}
\end{array},\\
\begin{array}{cccc}
\cline{1-4}
A_{1}&\cdots&A_{k-2}&A_{k-1}\\
\cline{4-4}
&&&A_{k-1}
\end{array}=
\begin{array}{rccc}
\cline{2-4}
s\paren{A_{k-2,k-1}A_{k-1}}&A_{1}&\cdots&A_{k-2}
\end{array},
\end{align}
where $A_{\alpha,\beta}$ is defined by 
\begin{align}
\bce{A_{\alpha},A_{\beta},A_{\alpha,\beta}}=\bce{X,Y,Z} \quad \text{for}~ A_{\alpha}\neq A_{\beta}.
\end{align}
In both cases, $(k-1,0)$ operators are generated.

$k$-support operators are only generated by
\begin{align}
\begin{array}{cccc}
\cline{1-4}
A_{1}&A_{2}&\cdots&A_{k-1}\\
\cline{1-1}
A_{1,2}
\end{array}=
\begin{array}{rcccc}
\cline{2-5}
s\paren{A_{1}A_{1,2}}&A_{2}&A_{2}&\cdots&A_{k-1}
\end{array},\\
\begin{array}{cccc}
\cline{1-4}
A_{1}&\cdots&A_{k-2}&A_{k-1}\\
\cline{4-4}
&&&A_{k-2,k-1}
\end{array}=
\begin{array}{rcccc}
\cline{2-5}
s\paren{A_{k-1}A_{k-2,k-1}}&A_{1}&\cdots&A_{k-2}&A_{k-2}
\end{array},
\end{align}
and if $A_{p+1}\neq A_{p-1}$ for $1< p < k-1$,
\begin{align}
\begin{array}{ccccccc}
\cline{1-7}
A_{1}&\cdots&A_{p-1}&A_{p}&A_{p+1}&\cdots&A_{k-1}\\
\cline{4-4}
&&&A_{p-1}&&&
\end{array}=
\begin{array}{rccccccc}
\cline{2-8}
s\paren{A_{p+1}A_{p-1}}&A_{1}&\cdots&A_{p-1}&A_{p+1}&A_{p+1}&\cdots&A_{k-1}
\end{array},\\
\begin{array}{ccccccc}
\cline{1-7}
A_{1}&\cdots&A_{p-1}&A_{p}&A_{p+1}&\cdots&A_{k-1}\\
\cline{4-4}
&&&A_{p+1}&&&
\end{array}=
\begin{array}{rccccccc}
\cline{2-8}
s\paren{A_{p-1}A_{p+1}}&A_{1}&\cdots&A_{p-1}&A_{p-1}&A_{p+1}&\cdots&A_{k-1}
\end{array}.
\end{align}
Therefore, only $(k,1)$ operators are generated.
Note that since $A_{p-1}\neq A_{p}$ and $A_{p}\neq A_{p+1}$, $A_{p+1}=A_{p-1,p}$ if $A_{p-1}\neq A_{p+1}$.

Here, let us consider $(k,1)$ operators generated by the commutators. In the case of $k>3$, $(k-1,1)$ operators are needed in $Q_{k}$ to cancel coefficients of them. To calculate coefficients of $(k-1,1)$ operators, we  first consider a $(k,1)$ operator whose hole is at the leftmost side, namely, $\overline{A_{1}^{2}A_{2}A_{3}\cdots A_{k-2}}$.  In the exceptional case of $k=3$, $(k-1,1)$ operators cannot exist because $(l, m)$ operators satisfy $l-m\geq 2$ by definition. In this case, $\overline{A_{1}A_{2}A_{3}\cdots A_{k-2}}=\overline{A_{1}^{2}}$, and its hole is also at the rightmost side, and the coefficients of $(3,1)$ operators are cancelled by only $(3,0)$ operators because
\begin{align}
&s\paren{A_{1}A_{1,2}}J_{A_{1,2}}q_{\overline{A_{1}A_{2}}}^{3,0}+s\paren{A_{1,2}A_{1}}J_{A_{1}}q_{\overline{A_{1,2}A_{2}}}^{3,0}+
s\paren{A_{1}A_{1,2}}J_{A_{1,2}}q_{\overline{A_{2}A_{1}}}^{3,0}+s\paren{A_{1,2}A_{1}}J_{A_{1}}q_{\overline{A_{2}A_{1,2}}}^{3,0}\nonumber\\
=~&J_{X}J_{Y}J_{Z}\bck{-r^{3.0}\paren{\overline{A_{1}A_{2}}}-r^{3.0}\paren{\overline{A_{1,2}A_{2}}}+r^{3.0}\paren{\overline{A_{2}A_{1}}}+r^{3.0}\paren{\overline{A_{2}A_{1,2}}}}=0
\end{align}
is satisfied. We assume $k>3$ hereafter. 
The only commutator of $(k-1,1)$ operator and $H$ that the $(k,1)$ operator generates is 
\begin{align}
\centering
\begin{array}{ccccc}
\cline{1-4}
A_{1}^{2}&A_{2}&\cdots&A_{k-3}&\\
\cline{5-5}
&&&&A_{k-2}
\end{array}=
\begin{array}{rccccc}
\cline{2-5}
s\paren{A_{k-3}A_{k-2}}&A_{1}^{2}&A_{2}&\cdots&A_{k-3}A_{k-2}.
\end{array}
\end{align}
Therefore, we have
\begin{align}
&s\paren{A_{k-3}A_{k-2}}J_{A_{k-2}}q^{k-1,1}_{\overline{A_{1}^{2}A_{2}\cdots A_{k-3}}}+s\paren{A_{2}A_{1,2}}J_{A_{1,2}}q^{k,0}_{\overline{A_{2}A_{1}A_{2}\cdots A_{k-2}}}
+s\paren{A_{1,2}A_{2}}J_{A_{2}}q^{k,0}_{\overline{A_{1,2}A_{1}A_{2}\cdots A_{k-2}}}
\nonumber\\
&+s\paren{A_{1}A_{2}}J_{A_{2}}q^{k,0}_{\overline{A_{1}A_{1,2}A_{2}\cdots A_{k-2}}}
=0.\label{supeq:(k,1)}
\end{align}
By using the function $r$ in 
Eq.~\eqref{supeq:defr},
Eq.~\eqref{supeq:(k,1)} becomes
\begin{align}
r^{k-1,1}\paren{\overline{A_{1}^{2}A_{2}\cdots A_{k-3}}}-r^{k,0}\paren{\overline{A_{2}A_{1}A_{2}\cdots A_{k-2}}}-r^{k,0}\paren{\overline{A_{1,2}A_{1}A_{2}\cdots A_{k-2}}}+r^{k,0}\paren{\overline{A_{1}A_{1,2}A_{2}\cdots A_{k-2}}}=0,
\end{align}
therefore, we obtain
\begin{align}
r^{k-1,1}\paren{\overline{A_{1}^{2}A_{2}\cdots A_{k-3}}}=1.
\label{supeq:(k-1,1)r1}
\end{align}
Here, we have used identities for $A_{\alpha}\neq A_{\beta}$:
\begin{align}
&s\paren{A_{\alpha}A_{\beta}A_{\alpha,\beta}}=1,\quad s\paren{A_{\alpha}A_{\beta}}=-s\paren{A_{\beta}A_{\alpha}},\quad s\paren{A_{\alpha}A_{\beta}}=-s\paren{A_{\alpha}A_{\alpha,\beta}},\\
&J_{A_{\alpha}}J_{A_{\beta}}J_{A_{\alpha,\beta}}=J_{X}J_{Y}J_{Z}.
\end{align}
In a similar manner, we have
\begin{align}
r^{k-1,1}\paren{\overline{A_{1}A_{2}\cdots A_{k-3}^{2}}}=1.
\label{supeq:(k-1,1)r2}
\end{align}

We next consider a $(k,1)$ operator $\overline{A_{1}\cdots A_{p-1}A_{p}^{2}A_{p+1}\cdots A_{k-2}}$ for $1<p<k-2$. In this case, there exist two $(k-1,1)$ operators which generate it:
\begin{align}
&\begin{array}{cccccccc}
\cline{2-8}
&A_{2}&\cdots&A_{p-1}&A_{p}^{2}&A_{p+1}&\cdots&A_{k-2}\\
\cline{1-1}
A_{1}&&&&&&&
\end{array}=
\begin{array}{rcccccccc}
\cline{2-9}
s\paren{A_{2}A_{1}}&A_{1}&A_{2}&\cdots&A_{p-1}&A_{p}^{2}&A_{p+1}&\cdots&A_{k-2}
\end{array},\\
&\begin{array}{cccccccc}
\cline{1-7}
A_{1}&\cdots&A_{p-1}&A_{p}^{2}&A_{p+1}&\cdots&A_{k-3}&\\
\cline{8-8}
&&&&&&&A_{k-2}
\end{array}=
\begin{array}{rcccccccc}
\cline{2-9}
s\paren{A_{k-3}A_{k-2}}&A_{1}&\cdots&A_{p-1}&A_{p}^{2}&A_{p+1}&\cdots&A_{k-3}&A_{k-2}
\end{array}.
\end{align}
Therefore, we obtain the condition to cancel the coefficient of $\overline{A_{1}\cdots A_{p-1}A_{p}^{2}A_{p+1}\cdots A_{k-2}}$:
\begin{align}
&r^{k-1,1}\paren{\overline{A_{1}\cdots A_{p-1}A_{p}^{2}A_{p+1}\cdots A_{k-3}}}
-r^{k-1,1}\paren{\overline{A_{2}\cdots A_{p-1}A_{p}^{2}A_{p+1}\cdots A_{k-2}}}\nonumber\\
&+r^{k,0}\paren{\overline{A_{1}\cdots A_{p-1}A_{p}A_{p, p+1}A_{p+1}\cdots A_{k-2}}}
-r^{k,0}\paren{\overline{A_{1}\cdots A_{p-1}A_{p-1,p}A_{p}A_{p+1}\cdots A_{k-2}}}=0,
\end{align}
and we obtain
\begin{align}
r^{k-1,1}\paren{\overline{A_{1}\cdots A_{p-1}A_{p}^{2}A_{p+1}\cdots A_{k-3}}}
=r^{k-1,1}\paren{\overline{A_{2}\cdots A_{p-1}A_{p}^{2}A_{p+1}\cdots A_{k-2}}}.
\label{supeq:(k-1,1)r3}
\end{align}
Combining Eq.~\eqref{supeq:(k-1,1)r1}, \eqref{supeq:(k-1,1)r2}, and \eqref{supeq:(k-1,1)r3}, we obtain all the coefficients of $(k-1,1)$ operators:
\begin{align}
r^{k-1,1}\paren{\overline{A_{1}\cdots A_{p-1}A_{p}^{2}A_{p+1}\cdots A_{k-3}}}=1 \quad\text{for}~1\leq p \leq k-3.
\end{align}

We prove by induction that all the coefficients of $(k-m,m)$ operators $(0\leq m\leq \lfloor k/2\rfloor -1)$ $\overline{A_{1}^{1+m_{1}}A_{2}^{1+m_{2}}\cdots A_{k-2m-1}^{1+m_{k-2m-1}}}$ satisfy
\begin{align}
r^{k-m,m}\paren{\overline{A_{1}^{1+m_{1}}A_{2}^{1+m_{2}}\cdots A_{k-2m-1}^{1+m_{k-2m-1}}}}=1.
\label{supeq:(k-m,m)r1}
\end{align}
Suppose that Eq.~\eqref{supeq:(k-m,m)r1} is satisfied for an $m=m^{\prime}<\lfloor k/2\rfloor -1$. 
By considering the cancellation of the coefficient of a $(k-m^{\prime},m^{\prime}+1)$ operator 
$\overline{A_{1}^{1+m_{1}}A_{2}^{1+m_{2}}\cdots A_{k-2m^{\prime}-3}^{1+m_{k-2m^{\prime}-3}} A_{k-2m^{\prime}-2}}$, where $m_{1}\geq 1$, we have
\begin{align}
&r^{k-m^{\prime}-1,m^{\prime}+1}\paren{\overline{A_{1}^{1+m_{1}}A_{2}^{1+m_{2}}\cdots A_{k-2m^{\prime}-3}^{1+m_{k-2m^{\prime}-3}}}}
-r^{k-m^{\prime},m^{\prime}}\paren{\overline{A_{2}A_{1}^{m_{1}}A_{2}^{1+m_{2}}\cdots A_{k-2m^{\prime}-2}}}\nonumber\\
&-r^{k-m^{\prime},m^{\prime}}\paren{\overline{A_{1,2}A_{1}^{m_{1}}A_{2}^{1+m_{2}}\cdots A_{k-2m^{\prime}-2}}}+r^{k-m^{\prime},m^{\prime}}\paren{\overline{A_{1}^{m_{1}}A_{1,2}A_{2}^{1+m_{2}}\cdots A_{k-2m^{\prime}-2}}}\nonumber\\
&+\sum_{\substack{p=2\\m_{p}\geq 1}}^{k-2m^{\prime}-3}\left[r^{k-m^{\prime},m^{\prime}}\paren{\overline{A_{1}^{1+m_{1}}\cdots A_{p}^{m_{p}}A_{p, p+1}A_{p+1}^{1+m_{p+1}}\cdots A_{k-2m^{\prime}-2}}}\right.
\nonumber\\
&-\left. r^{k-m^{\prime},m^{\prime}}\paren{\overline{A_{1}^{1+m_{1}}\cdots A_{p-1}^{1+m_{p-1}}A_{p-1, p}A_{p}^{m_{p}}\cdots A_{k-2m^{\prime}-2}}}\right]
\nonumber\\
&=0,
\end{align}
and by using the supposition, we obtain
\begin{align}
r^{k-m^{\prime}-1,m^{\prime}+1}\paren{\overline{A_{1}^{1+m_{1}}A_{2}^{1+m_{2}}\cdots A_{k-2m^{\prime}-3}^{1+m_{k-2m^{\prime}-3}}}}=1 \quad \text{for}~m_{1}\geq 1.
\label{supeq:(k-m,m)r2}
\end{align}
In a similar manner, by considering a $(k-m^{\prime},m^{\prime}+1)$ operator 
$\overline{A_{0}A_{1}^{1+m_{1}}\cdots A_{k-2m^{\prime}-3}^{1+m_{k-2m^{\prime}-3}}}$, where $m_{k-2m^{\prime}-3}\geq 1$,
\begin{align}
r^{k-m^{\prime}-1,m^{\prime}+1}\paren{\overline{A_{1}^{1+m_{1}}A_{2}^{1+m_{2}}\cdots A_{k-2m^{\prime}-3}^{1+m_{k-2m^{\prime}-3}}}}=1 \quad \text{for}~m_{k-2m^{\prime}-3}\geq 1,
\label{supeq:(k-m,m)r3}
\end{align}
is obtained. We next consider the cancellation of the coefficient of a $(k-m^{\prime},m^{\prime}+1)$ operator $\overline{A_{1}A_{2}^{1+m_{2}}\cdots A_{k-2m^{\prime}-3}^{1+m_{k-2m^{\prime}-3}} A_{k-2m^{\prime}-2}}$, and we have 
\begin{align}
&r^{k-m^{\prime}-1,m^{\prime}+1}\paren{\overline{A_{1}A_{2}^{1+m_{2}}\cdots A_{k-2m^{\prime}-3}^{1+m_{k-2m^{\prime}-3}}}}
-r^{k-m^{\prime}-1,m^{\prime}+1}\paren{\overline{A_{2}^{1+m_{2}}\cdots A_{k-2m^{\prime}-3}^{1+m_{k-2m^{\prime}-3}}A_{k-2m^{\prime}-2}}}\nonumber\\
&+\sum_{\substack{p=2\\m_{p}\geq 1}}^{k-2m^{\prime}-3}
\left[r^{k-m^{\prime},m^{\prime}}\paren{\overline{A_{1}A_{2}^{1+m_{2}}\cdots A_{p}^{m_{p}}A_{p, p+1}A_{p+1}^{1+m_{p+1}}\cdots A_{k-2m^{\prime}-3}^{1+m_{k-2m^{\prime}-3}}A_{k-2m^{\prime}-2}}}\right .\nonumber\\
&\left .-r^{k-m^{\prime},m^{\prime}}\paren{\overline{A_{1}A_{2}^{1+m_{2}}\cdots A_{p-1}^{1+m_{p-1}}A_{p-1, p}A_{p}^{m_{p}}\cdots A_{k-2m^{\prime}-3}^{1+m_{k-2m^{\prime}-3}}A_{k-2m^{\prime}-2}}}\right]\nonumber\\
&=0,
\end{align}
and by using the supposition, we obtain
\begin{align}
r^{k-m^{\prime}-1,m^{\prime}+1}\paren{\overline{A_{1}A_{2}^{1+m_{2}}\cdots A_{k-2m^{\prime}-3}^{1+m_{k-2m^{\prime}-3}}}}
=r^{k-m^{\prime}-1,m^{\prime}+1}\paren{\overline{A_{2}^{1+m_{2}}\cdots A_{k-2m^{\prime}-3}^{1+m_{k-2m^{\prime}-3}}A_{k-2m^{\prime}-2}}}.
\label{supeq:(k-m,m)r4}
\end{align}
Combining Eqs.~\eqref{supeq:(k-m,m)r2}-\eqref{supeq:(k-m,m)r3}, and ~\eqref{supeq:(k-m,m)r4}, 
all the coefficients of $(k-m^{\prime}-1,m^{\prime}+1)$ operators are determined as
\begin{align}
r^{k-m^{\prime}-1,m^{\prime}+1}\paren{\overline{A_{1}^{1+m_{1}}A_{2}^{1+m_{2}}\cdots A_{k-2m^{\prime}-3}^{1+m_{k-2m^{\prime}-3}}}}=1.
\label{supeq:(k-m,m)r5}
\end{align}
Finally, in the case of $k-2m^{\prime}-4\geq 1$, there exists a consistency condition for the cancellation of the coefficient of a $(k-m^{\prime}-1,m^{\prime}+2)$ operator $\overline{A_{1}^{1+m_{1}}A_{2}^{1+m_{2}}\cdots A_{k-2m^{\prime}-4}^{1+m_{k-2m^{\prime}-4}}}$, where $m_{1}\geq 1$, and $m_{k-2m^{\prime}-4}\geq 1$. 
For convenience, we write commutators of $(l, m)$ operators and $H$ as $[(l, m),H]$.
Note that the operator cannot be generated by $[(k-m^{\prime}-2,m^{\prime}+2),H]$.
The condition is represented as
\begin{align}
&-r^{k-m^{\prime}-1,m^{\prime}+1}\paren{\overline{A_{2}A_{1}^{m_{1}}A_{2}^{1+m_{2}}\cdots A_{k-2m^{\prime}-4}^{1+m_{k-2m^{\prime}-4}}}}-r^{k-m^{\prime}-1,m^{\prime}+1}\paren{\overline{A_{1,2}A_{1}^{m_{1}}A_{2}^{1+m_{2}}\cdots A_{k-2m^{\prime}-4}^{1+m_{k-2m^{\prime}-4}}}}\nonumber\\
&+r^{k-m^{\prime}-1,m^{\prime}+1}\paren{\overline{A_{1}^{m_{1}}A_{1,2}A_{2}^{1+m_{2}}\cdots A_{k-2m^{\prime}-4}^{1+m_{k-2m^{\prime}-4}}}}
-r^{k-m^{\prime}-1,m^{\prime}+1}\paren{\overline{A_{1}^{1+m_{1}}\cdots A_{k-2m^{\prime}-5}^{1+m_{k-2m^{\prime}-5}}A_{{k-2m^{\prime}-5},k-2m^{\prime}-4}A_{k-2m^{\prime}-4}^{m_{k-2m^{\prime}-4}}}}\nonumber\\
&+r^{k-m^{\prime}-1,m^{\prime}+1}\paren{\overline{A_{1}^{1+m_{1}}\cdots A_{k-2m^{\prime}-5}^{1+m_{k-2m^{\prime}-5}}A_{k-2m^{\prime}-4}^{m_{k-2m^{\prime}-4}}A_{{k-2m^{\prime}-5},k-2m^{\prime}-4}}}\nonumber\\
&+r^{k-m^{\prime}-1,m^{\prime}+1}\paren{\overline{A_{1}^{1+m_{1}}\cdots A_{k-2m^{\prime}-5}^{1+m_{k-2m^{\prime}-5}}A_{k-2m^{\prime}-4}^{m_{k-2m^{\prime}-4}}A_{{k-2m^{\prime}-5}}}}
\nonumber\\
&+\sum_{\substack{p=2\\m_{p}\geq 1}}^{k-2m^{\prime}-5}\left [r^{k-m^{\prime}-1,m^{\prime}+1}\paren{\overline{A_{1}^{1+m_{1}}\cdots A_{p}^{m_{p}}A_{p, p+1}A_{p+1}^{1+m_{p+1}}\cdots A_{k-2m^{\prime}-2}}}\right.\nonumber\\
&\left. -r^{k-m^{\prime}-1,m^{\prime}+1}\paren{\overline{A_{1}^{1+m_{1}}\cdots A_{p-1}^{1+m_{p-1}}A_{p-1, p}A_{p}^{m_{p}}\cdots A_{k-2m^{\prime}-2}}}\right]\nonumber\\
&=0,
\end{align}
and from Eq.~\eqref{supeq:(k-m,m)r5}, it is satisfied, and therefore, Eq.~\eqref{supeq:(k-m,m)r1} is proved. 

\subsection{B. Consistency condition for structure of $Q_{k}$}
In the previous subsection, we obtain the coefficients of $(k-m, m)$ operators. We next consider $[(k-m, m),H]$. In addition to $(k-m, m\pm 1)$ operators, operators which are not included in $(l, m)$ operators such as
\begin{align}
&\begin{array}{cccccccc}
\cline{1-8}
A_{1}^{1+m_{1}}&\cdots& A_{p-1}^{1+m_{p-1}}&A_{p}^{m_{p}}&A_{p}&A_{p+1}^{1+m_{p+1}}&\cdots &A_{k-2m-1}^{1+m_{k-2m-1}}\\
\cline{5-5}
&&&&A_{p}&&&
\end{array} \quad \text{for}~1\leq p <k-2m-1,
\label{supeq:i1}\\\nonumber\\
&\begin{array}{cccccccc}
\cline{1-8}
A_{1}^{1+m_{1}}&\cdots& A_{p-1}^{1+m_{p-1}}&A_{p}&A_{p}^{m_{p}}&A_{p+1}^{1+m_{p+1}}&\cdots &A_{k-2m-1}^{1+m_{k-2m-1}}\\
\cline{4-4}
&&&A_{p}&&&&
\end{array} \quad \text{for}~1< p \leq k-2m-1,
\label{supeq:i2}
\end{align}
can be generated if $m_{p}\geq 1$. However, one can prove that all the coefficients of these operators are zero by using Eq.~\eqref{supeq:(k-m,m)r1}. Here, we derive a condition of the cancellation for general $(l,m)$ operators and prove that Eq.~\eqref{supeq:(k-m,m)r1} satisfies the condition.

Let $\overline{A_{1}^{1+m_{1}}A_{2}^{1+m_{2}}\cdots A_{p}^{1+m_{p}}\cdots A_{l-m-1}^{1+m_{l-m-1}}}$ be an $(l,m)$ operator, where $m_{p}\geq 1$ and $1\leq p <l-m-1$. The commutator of it and $H$ generates an operator such as Eq.~\eqref{supeq:i1}:
\begin{align}
&\begin{array}{cccccccc}
\cline{1-8}
A_{1}^{1+m_{1}}&\cdots& A_{p-1}^{1+m_{p-1}}&A_{p}^{m_{p}}&I&A_{p+1}^{1+m_{p+1}}&\cdots &A_{l-m-1}^{1+m_{l-m-1}}
\end{array}
\nonumber\\\nonumber\\
\equiv~&
s\paren{A_{p,p+1}A_{p}}\times
\begin{array}{cccccccc}
\cline{1-8}
A_{1}^{1+m_{1}}&\cdots& A_{p-1}^{1+m_{p-1}}&A_{p}^{m_{p}}&A_{p}&A_{p+1}^{1+m_{p+1}}&\cdots &A_{l-m-1}^{1+m_{l-m-1}}\\
\cline{5-5}
&&&&A_{p}&&&
\end{array}.
\end{align}
This operator is also generated in a similar manner of Eq.~\eqref{supeq:i2} as 
\begin{align}
\begin{array}{cccccccc}
\cline{1-8}
A_{1}^{1+m_{1}}&\cdots& A_{p-1}^{1+m_{p-1}}&A_{p}^{m_{p}}&A_{p+1}&A_{p+1}^{1+m_{p+1}}&\cdots &A_{l-m-1}^{1+m_{l-m-1}}\\
\cline{5-5}
&&&&A_{p+1}&&&
\end{array}.
\end{align}
Therefore, we obtain the condition for these terms to be cancelled using the function $r$:
\begin{align}
r^{l,m}\paren{\overline{A_{1}^{1+m_{1}}\cdots A_{p-1}^{1+m_{p-1}}A_{p}^{1+m_{p}}A_{p+1}^{1+m_{p+1}}\cdots A_{l-m-1}^{1+m_{l-m-1}}}}=r^{l,m}\paren{\overline{A_{1}^{1+m_{1}}\cdots A_{p-1}^{1+m_{p-1}}A_{p}^{m_{p}}A_{p+1}^{2+m_{p+1}}\cdots A_{l-m-1}^{1+m_{l-m-1}}}},
\label{supeq:condition}
\end{align}
for all $1\leq p <l-m-1$. 

Obviously, Eq.~\eqref{supeq:(k-m,m)r1} satisfies Eq.~\eqref{supeq:condition}. 
In addition, all the coefficients we obtain below, \textit{i.e.}, 
Eqs.~(15)-(20),
also satisfy the condition, and therefore, only $(l, m)$ operators are included in $Q_{k}$.

\subsection{C. Conditions for coefficients of $Q_{k}$}
We derive conditions for coefficients of $(k-2n-m,m)$ operators for $n>0$. Suppose that, for all $0\leq n^{\prime}<n$ and $0\leq m^{\prime}\leq k/2 -n^{\prime}-1$, all the coefficients of $(k-2n^{\prime}-m^{\prime},m^{\prime})$ operators are determined. We can add $(k-2n+1,0)$ operators to $Q_{k}$ if $(k-2n+2,0)$ operators generated by their commutators with $H$ are cancelled. This degree of freedom corresponds to the addition of $Q_{k-2n+1}$ to $Q_{k}$. Here, we set coefficients of $(k-2n+1,0)$ operators zero. To obtain coefficients of $(k-2n,0)$ operators, we next consider $(k-2n+1,0)$ operators generated by commutators. Let $\overline{A_{1}A_{2}\cdots A_{k-2n}}$ be a $(k-2n+1,0)$ operator. The condition for coefficients of them to be cancelled is given as
\begin{align}
&r^{k-2n,0}\paren{\overline{A_{1}A_{2}\cdots A_{k-2n-1}}}-r^{k-2n,0}\paren{\overline{A_{2}A_{3}\cdots A_{k-2n}}}\nonumber\\
+~&J^{2}_{A_{1,2}}r^{k-2n+1,1}\paren{\overline{A_{2}A_{2}A_{3}\cdots A_{k-2n}}}
-J^{2}_{A_{k-2n-1,k-2n}}r^{k-2n+1,1}\paren{\overline{A_{1}\cdots A_{k-2n-2}A_{k-2n-1}A_{k-2n-1}}}\nonumber\\
+~&\sum_{p=2}^{k-2n-1}\bck{J^{2}_{A_{p-1}}r^{k-2n-1,1~\text{or}~2}\paren{\overline{A_{1}\cdots A_{p-1}A_{p+1}A_{p+1}\cdots A_{k-2n}}}-J^{2}_{A_{p+1}}r^{k-2n-1,1~\text{or}~2}\paren{\overline{A_{1}\cdots A_{p-1}A_{p-1}A_{p+1}\cdots A_{k-2n}}}}
\nonumber\\
+~&J^{2}_{A_{2}}r^{k-2n+2,0}\paren{\overline{A_{2}A_{1}A_{2}\cdots A_{k-2n}}}
+J^{2}_{A_{1,2}}r^{k-2n+2,0}\paren{\overline{A_{1,2}A_{1}A_{2}\cdots A_{k-2n}}}
\nonumber\\
-~&J^{2}_{A_{k-2n-1}}r^{k-2n+2,0}\paren{\overline{A_{1}\cdots A_{k-2n-1}A_{k-2n}A_{k-2n-1}}}
-J^{2}_{A_{k-2n-1,k-2n}}r^{k-2n+2,0}\paren{\overline{A_{1}\cdots A_{k-2n-1}A_{k-2n}A_{k-2n-1,k-2n}}}
\nonumber\\
=~&0.
\label{supeq:condition0hole}
\end{align}
In the third line, $r^{k-2n-1,1~\text{or}~2}=r^{k-2n-1,1}$ if $A_{p-1}\neq A_{p+1}$, and $r^{k-2n-1,2}$ if $A_{p-1}= A_{p+1}$. However, Eq.~\eqref{supeq:condition0hole} does not depend on the function $r^{k-2n-1,2}$ because if $A_{p-1}=A_{p+1}$, the sum of two terms of $p$ in the third line is zero.
Eq.~\eqref{supeq:condition0hole} is invariant under the transformation $r^{k-2n,0}\to r^{k-2n,0}+a$, where $a$ is an arbitrary constant. This corresponds to the addition of $a Q_{k-2n}$ to $Q_{k}$, and we can fix this degree of freedom freely. In this Letter, we fix it for the coefficients of $S_{0}\equiv 1$ to be zero. After this fixing, the function $r^{k-2n,0}$ is uniquely determined.

We further suppose that, for $0\leq m^{\prime\prime}\leq m-1$, all the coefficients of $(k-2n-m^{\prime\prime},m^{\prime\prime})$ operators are determined. We derive conditions for coefficients of $(k-2n-m,m)$ operators for $n>0$ and $m>0$. We consider $(k-2n-m+1,m)$ operators generated by commutators. Let $\overline{A_{1}^{1+m_{1}}A_{2}^{1+m_{2}}\cdots A_{k-2n-2m}^{1+m_{k-2n-2m}}}$ be a $(k-2n-m+1,m)$ operator. 
We first consider the case of $m_{1}\geq 1$ and $m_{k-2n-2m}=0$. In this case, conditions of the cancellation are given as
\begin{align}
&r^{\wsup,m}\paren{\overline{A_{1}^{1+m_{1}}A_{2}^{1+m_{2}}\cdots A_{\knm}^{1+m_{\knm}}}}
-r^{\wsup+1,m-1}\paren{\overline{A_{2}A_{1}^{m_{1}}A_{2}^{1+m_{2}}\cdots A_{\knm}^{1+m_{\knm}}A_{\knm+1}}}
\nonumber\\
-~&r^{\wsup+1,m-1}\paren{\overline{A_{1,2}A_{1}^{m_{1}}A_{2}^{1+m_{2}}\cdots A_{\knm}^{1+m_{\knm}}A_{\knm+1}}}
+r^{\wsup+1,m-1}\paren{\overline{A_{1}^{m_{1}}A_{1,2}A_{2}^{1+m_{2}}\cdots A_{\knm}^{1+m_{\knm}}A_{\knm+1}}}
\nonumber\\
+~&\sum_{\substack{p=2\\m_{p}\geq 1}}^{\knm}
\left[r^{\wsup+1,m-1}\paren{\overline{A_{1}^{1+m_{1}}\cdots A_{p}^{m_{p}}A_{p, p+1} A_{p+1}^{1+m_{p+1}}\cdots A_{\knm}^{1+m_{\knm}}A_{\knm+1}}}\right .
\nonumber\\
-~&\left. r^{\wsup+1,m-1}\paren{\overline{A_{1}^{1+m_{1}}\cdots A_{p-1}^{1+m_{p-1}}A_{p-1,p}A_{p}^{m_{p}}\cdots A_{\knm}^{1+m_{\knm}}A_{\knm+1}}}\right]
\nonumber\\
+~&\sum_{\substack{p=2\\m_{p}=0}}^{\knm}
\left[J^{2}_{A_{p-1}}r^{\wsup+1,m+1~\text{or}~m+2}\paren{\overline{A_{1}^{1+m_{1}}\cdots
A_{p-1}^{1+m_{p-1}}A_{p+1}^{2+m_{p+1}}A_{p+2}^{1+m_{p+2}}\cdots A_{\knm}^{1+m_{\knm}}A_{\knm+1}}}\right.
\nonumber\\
-~&\left. J^{2}_{A_{p+1}}r^{\wsup+1,m+1~\text{or}~m+2}\paren{\overline{A_{1}^{1+m_{1}}\cdots
A_{p-2}^{1+m_{p-2}}A_{p-1}^{2+m_{p-1}}A_{p+1}^{1+m_{p+1}}\cdots A_{\knm}^{1+m_{\knm}}A_{\knm+1}}}\right ]
\nonumber\\
-~&J^{2}_{A_{\knm,\knm+1}}r^{\wsup+1,m+1}\paren{\overline{A_{1}^{1+m_{1}}\cdots A_{\knm-1}^{1+m_{\knm-1}}A_{\knm}^{2+m_{\knm}}}}
\nonumber\\
+~&J^{2}_{A_{2}}r^{\wsup+2,m}\paren{\overline{A_{2}A_{1}^{1+m_{1}}A_{2}^{1+m_{2}}\cdots A_{\knm}^{1+m_{\knm}}A_{\knm+1}}}
+J^{2}_{A_{1,2}}r^{\wsup+2,m}\paren{\overline{A_{1,2}A_{1}^{1+m_{1}}A_{2}^{1+m_{2}}\cdots A_{\knm}^{1+m_{\knm}}A_{\knm+1}}}
\nonumber\\
-~&J^{2}_{A_{\knm}}r^{\wsup+2,m}\paren{\overline{A_{1}^{1+m_{1}}A_{2}^{1+m_{2}}\cdots A_{\knm}^{1+m_{\knm}}A_{\knm+1}A_{\knm}}}
-J^{2}_{A_{\knm,\knm+1}}r^{\wsup+2,m}\paren{\overline{A_{1}^{1+m_{1}}A_{2}^{1+m_{2}}\cdots A_{\knm}^{1+m_{\knm}}A_{\knm+1}A_{\knm,\knm+1}}}
\nonumber\\
=~&0,
\label{supeq:conditionmhole1}
\end{align}
where we have used $\wsup=k-2n-m$ and $\knm=k-2n-2m-1$ defined in Eq.~\eqref{supeq:knmdef}.

In the case of $m_{1}=0$ and $m_{k-2n-2m}=0$, conditions are given as
\begin{align}
&r^{\wsup,m}\paren{\overline{A_{1}A_{2}^{1+m_{2}}\cdots A_{\knm}^{1+m_{\knm}}}}
-r^{\wsup,m}\paren{\overline{A_{2}^{1+m_{2}}\cdots A_{\knm}^{1+m_{\knm}}A_{\wsup}}}
\nonumber\\
+~&\sum_{\substack{p=2\\m_{p}\geq 1}}^{\knm}
\left[r^{\wsup+1,m-1}\paren{\overline{A_{1}A_{2}^{1+m_{2}}\cdots A_{p}^{m_{p}}A_{p, p+1} A_{p+1}^{1+m_{p+1}}\cdots A_{\knm}^{1+m_{\knm}}A_{\knm+1}}}\right .
\nonumber\\
-~&\left. r^{\wsup+1,m-1}\paren{\overline{A_{1}A_{2}^{1+m_{2}}\cdots A_{p-1}^{1+m_{p-1}}A_{p-1,p}A_{p}^{m_{p}}\cdots A_{\knm}^{1+m_{\knm}}A_{\knm+1}}}\right]
\nonumber\\
+~&J^{2}_{A_{1,2}}r^{\wsup+1,m+1}\paren{\overline{A_{2}^{2+m_{2}}A_{3}^{1+m_{3}}\cdots
 A_{\knm}^{1+m_{\knm}}A_{\knm+1}}}
 -J^{2}_{A_{\knm,\knm+1}}r^{\wsup+1,m+1}\paren{\overline{A_{1}A_{2}^{1+m_{2}}A_{3}^{1+m_{3}}\cdots
 A_{\knm-1}^{1+m_{\knm-1}}A_{\knm}^{2+m_{\knm}}}}
 \nonumber\\
+~&\sum_{\substack{p=2\\m_{p}=0}}^{\knm}
\left[J^{2}_{A_{p-1}}r^{\wsup+1,m+1~\text{or}~m+2}\paren{\overline{A_{1}A_{2}^{1+m_{2}}\cdots
A_{p-1}^{1+m_{p-1}}A_{p+1}^{2+m_{p+1}}A_{p+2}^{1+m_{p+2}}\cdots A_{\knm}^{1+m_{\knm}}A_{\knm+1}}}\right.
\nonumber\\
-~&\left. J^{2}_{A_{p+1}}r^{\wsup+1,m+1~\text{or}~m+2}\paren{\overline{A_{1}A_{2}^{1+m_{2}}\cdots
A_{p-2}^{1+m_{p-2}}A_{p-1}^{2+m_{p-1}}A_{p+1}^{1+m_{p+1}}\cdots A_{\knm}^{1+m_{\knm}}A_{\knm+1}}}\right ]
\nonumber\\
+~&J^{2}_{A_{2}}r^{\wsup+2,m}\paren{\overline{A_{2}A_{1}A_{2}^{1+m_{2}}\cdots A_{\knm}^{1+m_{\knm}}A_{\knm+1}}}
+J^{2}_{A_{1,2}}r^{\wsup+2,m}\paren{\overline{A_{1,2}A_{1}A_{2}^{1+m_{2}}\cdots A_{\knm}^{1+m_{\knm}}A_{\knm+1}}}
\nonumber\\
-~&J^{2}_{A_{\knm}}r^{\wsup+2,m}\paren{\overline{A_{1}A_{2}^{1+m_{2}}\cdots A_{\knm}^{1+m_{\knm}}A_{\knm+1}A_{\knm}}}
-J^{2}_{A_{\knm,\knm+1}}r^{\wsup+2,m}\paren{\overline{A_{1}A_{2}^{1+m_{2}}\cdots A_{\knm}^{1+m_{\knm}}A_{\knm+1}A_{\knm,\knm+1}}}
\nonumber\\
=~&0,
\label{supeq:conditionmhole2}
\end{align}
which relates two coefficients in the first line.
By using Eqs.~\eqref{supeq:conditionmhole1}-\eqref{supeq:conditionmhole2}, all the coefficients of $(k-2n-2m,m)$ operators are determined.

Consistency conditions are as follows. First, we consider the case of $m_{1}=0$ and $m_{k-2n-2m}\geq 1$. Conditions for the case are similar to Eq.~\eqref{supeq:conditionmhole1} and given as
\begin{align}
-~&r^{\wsup,m}\paren{\overline{A_{2}^{1+m_{2}}\cdots A_{\knm}^{1+m_{\knm}}A_{\knm+1}^{1+m_{\knm+1}}}}
+r^{\wsup+1,m-1}\paren{\overline{A_{1}A_{2}^{1+m_{2}}\cdots A_{\knm}^{1+m_{\knm}}A_{\knm+1}^{m_{\knm+1}}A_{\knm}}}
\nonumber\\
+~&r^{\wsup+1,m-1}\paren{\overline{A_{1}A_{2}^{1+m_{2}}\cdots A_{\knm}^{1+m_{\knm}}A_{\knm+1}^{m_{\knm+1}}A_{\knm,\knm+1}}}
-r^{\wsup+1,m-1}\paren{\overline{A_{1}A_{2}^{1+m_{2}}\cdots A_{\knm}^{1+m_{\knm}}A_{\knm,\knm+1}A_{\knm+1}^{m_{\knm+1}}}}
\nonumber\\
+~&\sum_{\substack{p=2\\m_{p}\geq 1}}^{\knm}
\left[r^{\wsup+1,m-1}\paren{\overline{A_{1}A_{2}^{1+m_{2}}\cdots A_{p}^{m_{p}}A_{p, p+1} A_{p+1}^{1+m_{p+1}}\cdots A_{\knm+1}^{1+m_{\knm+1}}}}\right .
\nonumber\\
-~&\left. r^{\wsup+1,m-1}\paren{\overline{A_{1}A_{2}^{1+m_{2}}\cdots A_{p-1}^{1+m_{p-1}}A_{p-1,p}A_{p}^{m_{p}}\cdots A_{\knm+1}^{1+m_{\knm+1}}}}\right]
\nonumber\\
+~&\sum_{\substack{p=2\\m_{p}=0}}^{\knm}
\left[J^{2}_{A_{p-1}}r^{\wsup+1,m+1~\text{or}~m+2}\paren{\overline{A_{1}A_{2}^{1+m_{2}}\cdots
A_{p-1}^{1+m_{p-1}}A_{p+1}^{2+m_{p+1}}A_{p+2}^{1+m_{p+2}}\cdots A_{\knm+1}^{1+m_{\knm+1}}}}\right.
\nonumber\\
-~&\left. J^{2}_{A_{p+1}}r^{\wsup+1,m+1~\text{or}~m+2}\paren{\overline{A_{1}A_{2}^{1+m_{2}}\cdots
A_{p-2}^{1+m_{p-2}}A_{p-1}^{2+m_{p-1}}A_{p+1}^{1+m_{p+1}}\cdots A_{\knm+1}^{1+m_{\knm+1}}}}\right ]
\nonumber\\
+~&J^{2}_{A_{1,2}}r^{\wsup+1,m+1}\paren{\overline{A_{2}^{2+m_{2}}A_{3}^{1+m_{3}}\cdots A_{\knm}^{1+m_{\knm}}A_{\knm+1}^{1+m_{\knm+1}}}}
\nonumber\\
-~&J^{2}_{A_{\knm}}r^{\wsup+2,m}\paren{\overline{A_{1}A_{2}^{1+m_{2}}\cdots A_{\knm}^{1+m_{\knm}}A_{\knm+1}^{1+m_{\knm+1}}A_{\knm}}}
-J^{2}_{A_{\knm,\knm+1}}r^{\wsup+2,m}\paren{\overline{A_{1}A_{2}^{1+m_{2}}\cdots A_{\knm}^{1+m_{\knm}}A_{\knm+1}^{1+m_{\knm+1}}A_{\knm,\knm+1}}}
\nonumber\\
+~&J^{2}_{A_{2}}r^{\wsup+2,m}\paren{\overline{A_{2}A_{1}A_{2}^{1+m_{2}}\cdots A_{\knm}^{1+m_{\knm}}A_{\knm+1}^{1+m_{\knm+1}}}}
+J^{2}_{A_{1,2}}r^{\wsup+2,m}\paren{\overline{A_{1,2}A_{1}A_{2}^{1+m_{2}}\cdots A_{\knm}^{1+m_{\knm}}A_{\knm+1}^{1+m_{\knm+1}}}}
\nonumber\\
=~&0.
\label{supeq:conditionmhole3}
\end{align}
Second, in the case of $k-2n-2m-2\geq 1$, by considering the cancellation of the coefficient of a $(k-2n-m, m+1)$ operator $\overline{A_{1}^{1+m_{1}}A_{2}^{1+m_{2}}\cdots A_{k-2n-2m-2}^{1+m_{k-2n-2m-2}}}$, where $m_{1}\geq 1$ and $m_{k-2n-2m-2}\geq 1$, we obtain
\begin{align}
-~&r^{\wsup,m}\paren{\overline{A_{2}A_{1}^{m_{1}}A_{2}^{1+m_{2}}\cdots A_{\knm-1}^{1+m_{\knm-1}}}}-r^{\wsup,m}\paren{\overline{A_{1,2}A_{1}^{m_{1}}A_{2}^{1+m_{2}}\cdots A_{\knm-1}^{1+m_{\knm-1}}}}
\nonumber\\
+~&r^{\wsup,m}\paren{\overline{A_{1}^{m_{1}}A_{1,2}A_{2}^{1+m_{2}}\cdots A_{\knm-1}^{1+m_{\knm-1}}}}
\nonumber\\
+~&r^{\wsup,m}\paren{\overline{A_{1}^{1+m_{1}}A_{2}^{1+m_{2}}\cdots A_{\knm-2}^{1+m_{\knm-2}}A_{\knm-1}^{m_{\knm-1}}A_{\knm-2}}}
+r^{\wsup,m}\paren{\overline{A_{1}^{1+m_{1}}A_{2}^{1+m_{2}}\cdots A_{\knm-2}^{1+m_{\knm-2}}A_{\knm-1}^{m_{\knm-1}}A_{\knm-2,\knm-1}}}
\nonumber\\
-~&r^{\wsup,m}\paren{\overline{A_{1}^{1+m_{1}}A_{2}^{1+m_{2}}\cdots A_{\knm-2}^{1+m_{\knm-2}}A_{\knm-2,\knm-1}A_{\knm-1}^{m_{\knm-1}}}}
\nonumber\\
+~&\sum_{\substack{p=2\\m_{p}\geq 1}}^{\knm-2}
\left[r^{\wsup,m}\paren{\overline{A_{1}^{1+m_{1}}\cdots A_{p}^{m_{p}}A_{p, p+1} A_{p+1}^{1+m_{p+1}}\cdots A_{\knm-1}^{1+m_{\knm-1}}}}\right .
\nonumber\\
-~&\left. r^{\wsup,m}\paren{\overline{A_{1}^{1+m_{1}}\cdots A_{p-1}^{1+m_{p-1}}A_{p-1,p}A_{p}^{m_{p}}\cdots A_{\knm-1}^{1+m_{\knm-1}}}}\right]
\nonumber\\
+~&\sum_{\substack{p=2\\m_{p}=0}}^{\knm-2}
\left[J^{2}_{A_{p-1}}r^{\wsup,m+2~\text{or}~m+3}\paren{\overline{A_{1}^{1+m_{1}}\cdots
A_{p-1}^{1+m_{p-1}}A_{p+1}^{2+m_{p+1}}A_{p+2}^{1+m_{p+2}}\cdots A_{\knm-1}^{1+m_{\knm-1}}}}\right.
\nonumber\\
-~&\left. J^{2}_{A_{p+1}}r^{\wsup,m+2~\text{or}~m+3}\paren{\overline{A_{1}^{1+m_{1}}\cdots
A_{p-2}^{1+m_{p-2}}A_{p-1}^{2+m_{p-1}}A_{p+1}^{1+m_{p+1}}\cdots A_{\knm-1}^{1+m_{\knm-1}}}}\right ]
\nonumber\\
+~&J^{2}_{A_{2}}r^{\wsup+1,m+1}\paren{\overline{A_{2}A_{1}^{1+m_{1}}A_{2}^{1+m_{2}}\cdots A_{\knm-1}^{1+m_{\knm-1}}}}
+J^{2}_{A_{1,2}}r^{\wsup+1,m+1}\paren{\overline{A_{1,2}A_{1}^{1+m_{1}}A_{2}^{1+m_{2}}\cdots A_{\knm-1}^{1+m_{\knm-1}}}}
\nonumber\\
-~&J^{2}_{A_{\knm-2}}r^{\wsup+1,m+1}\paren{\overline{A_{1}^{1+m_{1}}A_{2}^{1+m_{2}}\cdots A_{\knm-1}^{1+m_{\knm-1}}A_{\knm-2}}}
-J^{2}_{A_{\knm-2,\knm-1}}r^{\wsup+1,m+1}\paren{\overline{A_{1}^{1+m_{1}}A_{2}^{1+m_{2}}\cdots A_{\knm-1}^{1+m_{\knm-1}}A_{\knm-2,\knm-1}}}
\nonumber\\
=~&0.
\label{supeq:conditionlrhole}
\end{align}

As a result, by solving
Eqs.~\eqref{supeq:condition}-\eqref{supeq:conditionlrhole},
one can obtain all the coefficients.

\subsection{D. Recursive way to construct coefficients of $Q_{k}$}
In the previous subsections, we derive the conditions for the coefficients. Here, we show that one can calculate them in a simple recursive way. We first present the way and prove that the coefficients calculated by it are the solution of 
Eqs.~\eqref{supeq:condition}-\eqref{supeq:conditionlrhole}.

Suppose that $r^{k-2n-m, m}$ does not depend on where holes are, \textit{i.e.}, $r^{l, m}$ can be written as
\begin{align}
r^{k-2n-m, m}\paren{\overline{A_{1}^{1+m_{1}}A_{2}^{1+m_{2}}\cdots A_{k-2n-2m-1}^{1+m_{k-2n-2m-1}}}}\equiv R^{k-2n-m, m}\paren{A_{1}A_{2}\cdots A_{k-2n-2m-1}},
\label{supeq:defrR}
\end{align}
We note that the function $R$ is unknown here.
From Eq.~\eqref{supeq:(k-m,m)r1}, in the case of $m=0$,
\begin{align}
R^{k-m, m}\paren{A_{1}A_{2}\cdots A_{k-2m-1}}=1=S_{0}.
\end{align}
$S_{p}$ is the function defined in 
Eq.~(17):
\begin{align}
S_{p}\paren{A_{1}A_{2}\cdots A_{l}}\equiv 
\begin{cases}
\sum_{1\leq j1\leq j2\leq\cdots\leq jp\leq l} J^{2}_{A_{j1}}J^{2}_{A_{j2}}\cdots J^{2}_{A_{jp}} & (p\geq 1),\\
1 & (p=0),
\end{cases}
\end{align}
where $A_{1}A_{2}\cdots A_{l}$ is a character string of length $l\geq 1$, and $A_{1}$, $A_{2}$, \ldots, $A_{l}$ take one of $\bce{X,Y,Z}$, respectively. By definition, $S_{p}$ is symmetric with respect to the exchange of the characters $A_{\alpha}\leftrightarrow A_{\beta}$. Therefore, $S_{p}$ is a symmetric polynomial in $J^{2}_{A_{1}},J^{2}_{A_{2}},\ldots , J^{2}_{A_{l}}$ and depends only on the number of $X$, $Y$, and $Z$ in $A_{1}A_{2}\cdots A_{l}$.
In addition, we introduce $a_{n}$ in 
Eq.~(18):
\begin{align}
a_{n}\equiv \frac{J^{2}_{X}(J^{2(n+2)}_{Y}-J^{2(n+2)}_{Z})+J^{2}_{Y}(J^{2(n+2)}_{Z}-J^{2(n+2)}_{X})+J^{2}_{Z}(J^{2(n+2)}_{X}-J^{2(n+2)}_{Y})}{(J^{2}_{X}-J^{2}_{Y})(J^{2}_{Y}-J^{2}_{Z})(J^{2}_{Z}-J^{2}_{X})}.
\end{align}
For example, $a_{-2}=a_{-1}=0$, $a_{0}=1$, $a_{1}=J^{2}_{X}+J^{2}_{Y}+J^{2}_{Z}$, and $a_{2}=J^{4}_{X}+J^{4}_{Y}+J^{4}_{Z}+J^{2}_{X}J^{2}_{Y}+J^{2}_{Y}J^{2}_{Z}+J^{2}_{Z}J^{2}_{X}$.
$a_{n}$ is characterized as follows. Let us consider the division of a monomial $u^{n+2}$ by $(u-J^{2}_{X})(u-J^{2}_{Y})(u-J^{2}_{Z})$. $a_{n}$ is the coefficient of $u^{2}$ in the remainder:
\begin{align}
u^{n+2}=(u-J^{2}_{X})(u-J^{2}_{Y})(u-J^{2}_{Z})(\text{a polynomial with respect to}~u)+
a_{n}u^{2}+(a_{n+1}-a_{1}a_{n})u+J^{2}_{X}J^{2}_{Y}J^{2}_{Z}a_{n-1},
\end{align}
and therefore, we obtain an identity
\begin{align}
J^{2(n+2)}_{A}=a_{n}J^{4}_{A}+(a_{n+1}-a_{1}a_{n})J^{2}_{A}+J^{2}_{X}J^{2}_{Y}J^{2}_{Z}a_{n-1}\quad \text{for}~ A=X, Y, \text{or} ~Z.
\label{supeq:anid}
\end{align}

\begin{figure}[tb]
\centering
\includegraphics[width=\columnwidth,clip]{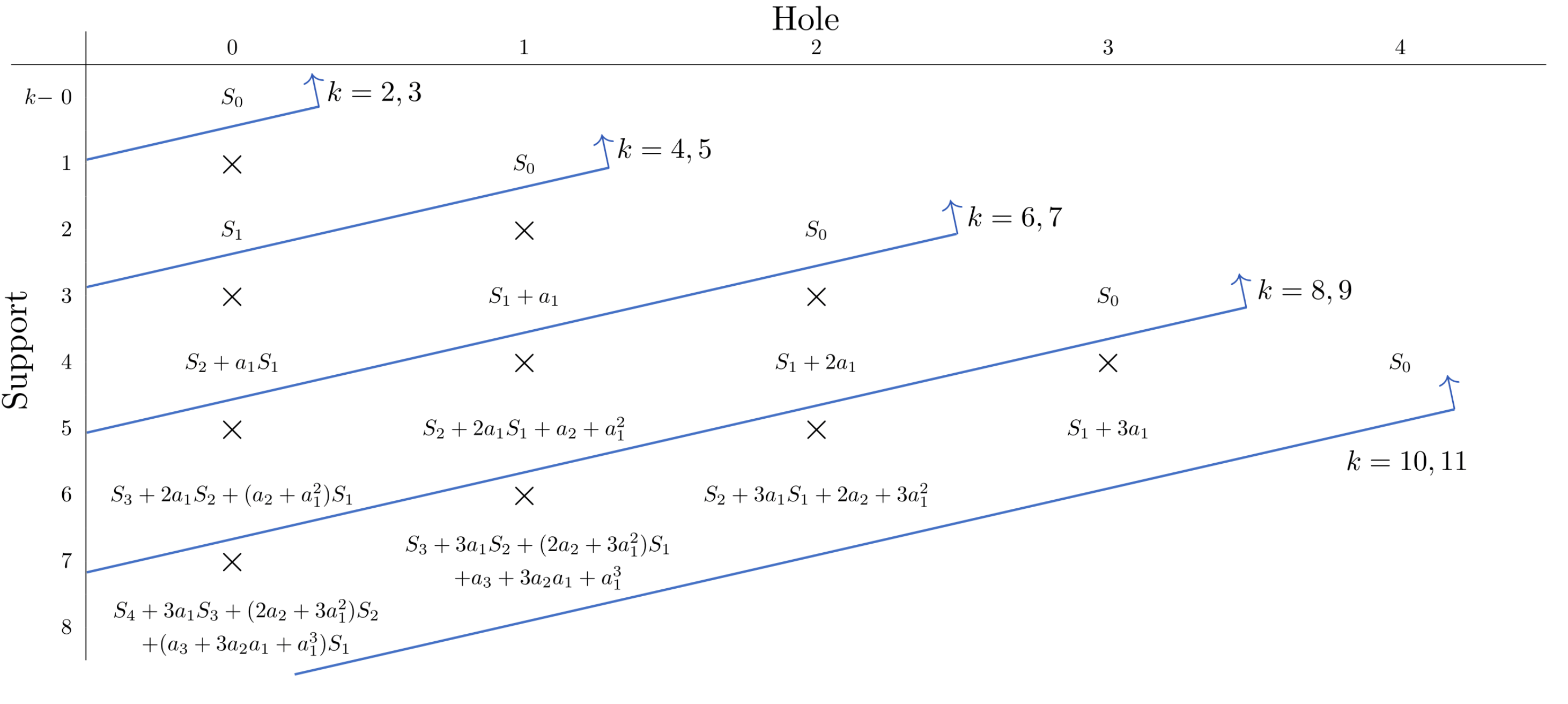}
\caption{$R^{k-2n-m,m}$ for $2\leq k\leq 11$, where the support is $k-2n-m$, and the hole is $m$.}
\label{supfig:rexample}
\end{figure}

After calculating $R^{k-2n-m^{\prime}, m^{\prime}}$ for all $0\leq m^{\prime}\leq k/2-n-1$, $R^{k-2(n+1)-m, m}$ for $0\leq m\leq k/2-(n+1)-1$
is obtained as follows. 
Suppose that $R^{k-2n-(m+1), m+1}$ calculated is written as
 \begin{align}
R^{k-2n-(m+1), m+1}\paren{A_{1}A_{2}\cdots A_{k-2n-2m-3}}=\sum_{\tilde{n}=0}^{n}g^{k-2n-(m+1), m+1}_{n-\tilde{n}}S_{\tilde{n}}\paren{A_{1}A_{2}\cdots A_{k-2n-2m-3}},
\label{supeq:rec1o}
\end{align}
where $g^{k-2n-(m+1), m+1}_{n+1-\tilde{n}}$ does not depend on $A_{1}A_{2}\cdots A_{k-2n-2m-3}$, and $g^{k-2n-(m+1), m+1}_{0}=1$.
Then 
$R^{k-2(n+1)-m, m}$ is obtained by the replacement $S_{\tilde{n}}\to S_{\tilde{n}+1}$ and the addition of $g^{k-2(n+1)-m, m}_{n+1}$:
\begin{align}
R^{k-2(n+1)-m, m}\paren{A_{1}A_{2}\cdots A_{k-2n-2m-3}}=&\sum_{\tilde{n}=0}^{n}g^{k-2n-(m+1), m+1}_{n-\tilde{n}}S_{\tilde{n}+1}\paren{A_{1}A_{2}\cdots A_{k-2n-2m-3}}
\label{supeq:rec1a}\nonumber\\
&+g^{k-2(n+1)-m, m}_{n+1}\\
\equiv&\sum_{\tilde{n}=0}^{n+1}g^{k-2(n+1)-m,m}_{n+1-\tilde{n}}S_{\tilde{n}}\paren{A_{1}A_{2}\cdots A_{k-2n-2m-3}}
\end{align}
where $g^{k-2(n+1)-m, m}_{\tilde{n}}$ is determined as
\begin{align}
g^{k-2(n+1)-m,m}_{\tilde{n}}&=g^{k-2n-(m+1),m+1}_{\tilde{n}}\quad \text{for}~0\leq \tilde{n}\leq n,
\label{supeq:rec1d}\\
g^{k-2(n+1), 0}_{n+1}&=0,\label{supeq:rec1b}\\
g^{k-2(n+1)-m, m}_{n+1}&=g^{k-2(n+1)-(m-1), m-1}_{n+1}+\sum_{\tilde{n}=0}^{n}~g^{k-2n-m, m}_{n-\tilde{n}}a_{\tilde{n}+1}\nonumber\\
&=\sum_{\tilde{n}=0}^{n+1}g^{k-2(n+1)-(m-1), m-1}_{n+1-\tilde{n}}a_{\tilde{n}}.
\label{supeq:rec1c}\end{align}
In this way, all the coefficients can be constructed. 
For example, Figure.~\ref{supfig:rexample} shows the function $R^{k-2n-m,m}$ for $2\leq k\leq 11$.

We prove that the coefficients satisfy Eqs.~\eqref{supeq:condition}-\eqref{supeq:conditionlrhole}. From Eq.~\eqref{supeq:defrR}, Eq.~\eqref{supeq:condition} is satisfied obviously. We next consider Eq.~\eqref{supeq:condition0hole}. It is useful for our proof to use properties of $S_{p}$ given as
\begin{align}
S_{p}\paren{A_{0}A_{1}A_{2}\cdots A_{l}}&=
\sum_{\tilde{p}=0}^{p}J^{2\tilde{p}}_{A_{0}}S_{p-\tilde{p}}\paren{A_{1}A_{2}\cdots A_{l}},
\label{supeq:sp1}\\
S_{p}\paren{A_{1}A_{2}\cdots A_{\tilde{l}-1}A_{\tilde{l}+1}\cdots A_{l}}&=S_{p}\paren{A_{1}A_{2}\cdots A_{\tilde{l}-1}A_{\tilde{l}}A_{\tilde{l}+1}\cdots A_{l}}-J^{2}_{A_{\tilde{l}}}S_{p-1}\paren{A_{1}A_{2}\cdots A_{\tilde{l}-1}A_{\tilde{l}}A_{\tilde{l}+1}\cdots A_{l}}.
\label{supeq:sp2}\end{align}
$R^{k-2(n-1),0}$ and $R^{k-2(n-1)-1,1}$ can be expressed as 
\begin{align}
R^{k-2(n-1),0}&=g_{0}S_{n-1}+g_{1}S_{n-2}+\cdots+g_{n-3}S_{2}+g_{n-2}S_{1}+g_{n-1}S_{0},
\label{supeq:k-2n+20}\\
R^{k-2(n-1)-1,1}&=h_{0}S_{n-1}+h_{1}S_{n-2}+\cdots+h_{n-3}S_{2}+h_{n-2}S_{1}+h_{n-1}S_{0},
\label{supeq:k-2n+11}
\end{align}
where $g_{0}=h_{0}=1$ and $g_{n-1}=0$.
Then, by using Eqs.~\eqref{supeq:rec1o}-\eqref{supeq:rec1c},
\begin{align}
R^{k-2n,0}&=h_{0}S_{n}+h_{1}S_{n-1}+\cdots+h_{n-3}S_{3}+h_{n-2}S_{2}+h_{n-1}S_{1},
\label{supeq:k-2n0}\\
h_{l}&=\sum_{\tilde{l}=0}^{l}g_{l-\tilde{l}}a_{\tilde{l}}
\label{supeq:hl}.
\end{align}
Substituting Eqs.~\eqref{supeq:k-2n+20}-\eqref{supeq:k-2n0} into the left-hand side of Eq.~\eqref{supeq:condition0hole} and using the properties Eqs.~\eqref{supeq:anid},~\eqref{supeq:sp1}-\eqref{supeq:sp2}, and \eqref{supeq:hl} with a notation $\overline{S}_{p}\equiv S_{p}\paren{A_{1}A_{2}\cdots A_{k-2n}}$, we obtain
\begin{align}
&\paren{J^{2}_{A_{1}}-J^{2}_{A_{k-2n}}}\sum_{\tilde{n}=0}^{n-1}h_{n-1-\tilde{n}}\overline{S}_{\tilde{n}}
\nonumber\\
+~&\paren{J^{2}_{A_{1,2}}-J^{2}_{A_{k-2n-1,k-2n}}}
\sum_{\tilde{n}=0}^{n-1}h_{n-1-\tilde{n}}\overline{S}_{\tilde{n}}
+\paren{J^{2}_{A_{k-2n-1,k-2n}}J^{2}_{A_{k-2n}}-J^{2}_{A_{1,2}}J^{2}_{A_{1}}}
\sum_{\tilde{n}=0}^{n-2}h_{n-2-\tilde{n}}\overline{S}_{\tilde{n}}
\nonumber\\
+~&\sum_{p=2}^{k-2n-1}\bck{\paren{J^{2}_{A_{p-1}}-J^{2}_{A_{p+1}}}\sum_{\tilde{n}=0}^{n-1}h_{n-1-\tilde{n}}\overline{S}_{\tilde{n}}+
\paren{J^{2}_{A_{p}}J^{2}_{A_{p+1}}-J^{2}_{A_{p-1}}J^{2}_{A_{p}}}\sum_{\tilde{n}=0}^{n-2}h_{n-2-\tilde{n}}\overline{S}_{\tilde{n}}}
\nonumber\\
+~&\sum_{\tilde{n}=0}^{n-1}\sum_{n_{1}=0}^{\tilde{n}}\paren{J^{2(n_{1}+1)}_{A_{2}}+J^{2(n_{1}+1)}_{A_{1,2}}-J^{2(n_{1}+1)}_{A_{k-2n-1}}-J^{2(n_{1}+1)}_{A_{k-2n-1,k-2n}}}g_{n-1-\tilde{n}}\overline{S}_{\tilde{n}-n_{1}}\nonumber\\
=~&\paren{J^{2}_{A_{1}}-J^{2}_{A_{k-2n}}}\sum_{\tilde{n}=0}^{n-1}h_{n-1-\tilde{n}}\overline{S}_{\tilde{n}}
\nonumber\\
+~&\paren{J^{2}_{A_{1,2}}-J^{2}_{A_{k-2n-1,k-2n}}}
\sum_{\tilde{n}=0}^{n-1}h_{n-1-\tilde{n}}\overline{S}_{\tilde{n}}
+\paren{J^{2}_{A_{k-2n-1,k-2n}}J^{2}_{A_{k-2n}}-J^{2}_{A_{1,2}}J^{2}_{A_{1}}}
\sum_{\tilde{n}=0}^{n-2}h_{n-2-\tilde{n}}\overline{S}_{\tilde{n}}
\nonumber\\
+~&\paren{J^{2}_{A_{1}}+J^{2}_{A_{2}}-J^{2}_{A_{k-2n-1}}-J^{2}_{A_{k-2n}}}\sum_{\tilde{n}=0}^{n-1}h_{n-1-\tilde{n}}\overline{S}_{\tilde{n}}+
\paren{J^{2}_{A_{k-2n-1}}J^{2}_{A_{k-2n}}-J^{2}_{A_{1}}J^{2}_{A_{2}}}\sum_{\tilde{n}=0}^{n-2}h_{n-2-\tilde{n}}\overline{S}_{\tilde{n}}
\nonumber\\
+~&\sum_{\tilde{n}=0}^{n-1}\sum_{n_{1}=0}^{\tilde{n}}\bck{a_{n_{1}}\paren{J^{2}_{A_{k-2n}}-J^{2}_{A_{1}}}+a_{n_{1}-1}J^{2}_{X}J^{2}_{Y}J^{2}_{Z}\paren{1/J^{2}_{A_{k-2n}}-1/J^{2}_{A_{1}}}}g_{n-1-\tilde{n}}\overline{S}_{\tilde{n}-n_{1}}
\nonumber\\
=~&\paren{J^{2}_{A_{1}}-J^{2}_{A_{k-2n}}}\sum_{\tilde{n}=0}^{n-1}h_{n-1-\tilde{n}}\overline{S}_{\tilde{n}}
\nonumber\\
+~&\paren{J^{2}_{A_{k-2n-1}}J^{2}_{A_{k-2n}}+J^{2}_{A_{k-2n-1,k-2n}}J^{2}_{A_{k-2n}}-J^{2}_{A_{1}}J^{2}_{A_{2}}-J^{2}_{A_{1}}J^{2}_{A_{1,2}}}
\sum_{\tilde{n}=0}^{n-2}h_{n-2-\tilde{n}}\overline{S}_{\tilde{n}}
\nonumber\\
+~&\sum_{\tilde{n}=0}^{n-1}\sum_{n_{1}=0}^{n-\tilde{n}-1}\bck{a_{n_{1}}\paren{J^{2}_{A_{k-2n}}-J^{2}_{A_{1}}}+a_{n_{1}-1}\paren{J^{2}_{A_{k-2n-1}}J^{2}_{A_{k-2n-1,k-2n}}-J^{2}_{A_{2}}J^{2}_{A_{1,2}}}}g_{n-n_{1}-\tilde{n}-1}\overline{S}_{\tilde{n}}
\nonumber\\
=~&\paren{J^{2}_{A_{1}}-J^{2}_{A_{k-2n}}}\sum_{\tilde{n}=0}^{n-1}h_{n-1-\tilde{n}}\overline{S}_{\tilde{n}}
\nonumber\\
+~&\paren{J^{2}_{A_{k-2n-1}}J^{2}_{A_{k-2n}}+J^{2}_{A_{k-2n-1,k-2n}}J^{2}_{A_{k-2n}}-J^{2}_{A_{1}}J^{2}_{A_{2}}-J^{2}_{A_{1}}J^{2}_{A_{1,2}}}
\sum_{\tilde{n}=0}^{n-2}h_{n-2-\tilde{n}}\overline{S}_{\tilde{n}}
\nonumber\\
+~&\paren{J^{2}_{A_{k-2n}}-J^{2}_{A_{1}}}\sum_{\tilde{n}=0}^{n-1}h_{n-1-\tilde{n}}\overline{S}_{\tilde{n}}+\paren{J^{2}_{A_{k-2n-1}}J^{2}_{A_{k-2n-1,k-2n}}-J^{2}_{A_{2}}J^{2}_{A_{1,2}}}\sum_{\tilde{n}=0}^{n-2}h_{n-2-\tilde{n}}\overline{S}_{\tilde{n}}
\nonumber\\
=~&\sum_{\tilde{n}=0}^{n-2}\paren{J^{2}_{X}J^{2}_{Y}+J^{2}_{Y}J^{2}_{Z}+J^{2}_{Z}J^{2}_{X}-J^{2}_{X}J^{2}_{Y}-J^{2}_{Y}J^{2}_{Z}-J^{2}_{Z}J^{2}_{X}}h_{n-2-\tilde{n}}\overline{S}_{\tilde{n}}=0,
\end{align}
therefore, Eq.~\eqref{supeq:condition0hole} is satisfied.

We prove that Eq.~\eqref{supeq:conditionmhole1} is satisfied. $R^{k-2(n-1)-m,m}$ and $R^{k-2(n-1)-(m+1),m+1}$ for $m\geq 1$ can be expressed as 
\begin{align}
R^{k-2(n-1)-m,m}&=g_{0}S_{n-1}+g_{1}S_{n-2}+\cdots+g_{n-3}S_{2}+g_{n-2}S_{1}+g_{n-1}S_{0},
\label{supeq:k-2(n-1)-mm}\\
R^{k-2(n-1)-(m+1),m+1}&=h_{0}S_{n-1}+h_{1}S_{n-2}+\cdots+h_{n-3}S_{2}+h_{n-2}S_{1}+h_{n-1}S_{0},
\label{supeq:k-2(n-1)-(m+1)m+1}
\end{align}
where $g_{0}=h_{0}=1$. $R^{k-2n-(m-1),m-1}$ and $R^{k-2n-m,m}$ are expressed as
\begin{align}
R^{k-2n-(m-1),m-1}&=g_{0}S_{n}+g_{1}S_{n-1}+\cdots+g_{n-3}S_{3}+g_{n-2}S_{2}+g_{n-1}S_{1}+g_{n},
\label{supeq:k-2n-(m-1)m-1}\\
R^{k-2n-m,m}&=h_{0}S_{n}+h_{1}S_{n-1}+\cdots+h_{n-3}S_{3}+h_{n-2}S_{2}+h_{n-1}S_{1}+h_{n},
\label{supeq:k-2n-mm}
\end{align}
Eq.~\eqref{supeq:hl} is satisfied also in this case.
Substituting Eqs.~\eqref{supeq:k-2(n-1)-mm}-\eqref{supeq:k-2n-mm} into the left-hand side of Eq.~\eqref{supeq:conditionmhole1} and using the properties Eqs.~\eqref{supeq:anid},~\eqref{supeq:sp1}-\eqref{supeq:sp2}, and \eqref{supeq:hl} with a notation $\overline{S}_{p}\equiv S_{p}\paren{A_{1}A_{2}\cdots A_{k-2n-2m}}$, we obtain
\begin{align}
&\sum_{\tilde{n}=0}^{n}h_{n-\tilde{n}}\overline{S}_{\tilde{n}}-J^{2}_{A_{\knm+1}}\sum_{\tilde{n}=0}^{n-1}h_{n-1-\tilde{n}}\overline{S}_{\tilde{n}}
-\sum_{\tilde{n}=0}^{n}\sum_{n_{1}=0}^{\tilde{n}}J^{2n_{1}}_{A_{2}}g_{n-\tilde{n}}\overline{S}_{\tilde{n}-n_{1}}
\nonumber\\
+~&\sum_{\substack{p=2\\m_{p}\geq 1}}^{\knm}\sum_{\tilde{n}=0}^{n}\sum_{n_{1}=0}^{\tilde{n}}\paren{J^{2n_{1}}_{A_{p, p+1}}-J^{2n_{1}}_{A_{p-1, p}}}g_{n-\tilde{n}}\overline{S}_{\tilde{n}-n_{1}}
\nonumber\\
+~&\sum_{\substack{p=2\\m_{p}=0}}^{\knm}\bck{\paren{J^{2}_{A_{p-1}}-J^{2}_{A_{p+1}}}\sum_{\tilde{n}=0}^{n-1}h_{n-1-\tilde{n}}\overline{S}_{\tilde{n}}+\paren{J^{2}_{A_{p}}J^{2}_{A_{p+1}}-J^{2}_{A_{p-1}}J^{2}_{A_{p}}}\sum_{\tilde{n}=0}^{n-2}h_{n-2-\tilde{n}}\overline{S}_{\tilde{n}}}
\nonumber\\
-~&J^{2}_{A_{\knm,\knm+1}}\sum_{\tilde{n}=0}^{n-1}h_{n-1-\tilde{n}}\overline{S}_{\tilde{n}}+J^{2}_{A_{\knm+1}}J^{2}_{A_{\knm,\knm+1}}\sum_{\tilde{n}=0}^{n-2}h_{n-2-\tilde{n}}\overline{S}_{\tilde{n}}
\nonumber\\
+~&\sum_{\tilde{n}=0}^{n-1}\sum_{n_{1}=0}^{\tilde{n}}\paren{J^{2(n_{1}+1)}_{A_{2}}+J^{2(n_{1}+1)}_{A_{1,2}}-J^{2(n_{1}+1)}_{A_{\knm}}-J^{2(n_{1}+1)}_{A_{\knm,\knm+1}}}g_{n-1-\tilde{n}}\overline{S}_{\tilde{n}-n_{1}}
\nonumber\\
=~&\sum_{\tilde{n}=0}^{n}h_{n-\tilde{n}}\overline{S}_{\tilde{n}}-J^{2}_{A_{\knm+1}}\sum_{\tilde{n}=0}^{n-1}h_{n-1-\tilde{n}}\overline{S}_{\tilde{n}}
-\sum_{\tilde{n}=0}^{n}\sum_{n_{1}=0}^{n-\tilde{n}}\paren{a_{n_{1}-1}J^{2}_{A_{2}}+a_{n_{1}}-a_{1}a_{n_{1}-1}+a_{n_{1}-2}J^{2}_{A_{1}}J^{2}_{A_{1,2}}}g_{n-n_{1}-\tilde{n}}\overline{S}_{\tilde{n}}
\nonumber\\
+~&\sum_{\substack{p=2\\m_{p}\geq 1}}^{\knm}\sum_{\tilde{n}=0}^{n}\sum_{n_{1}=0}^{n-\tilde{n}}\bck{a_{n_{1}-1}\paren{J^{2}_{A_{p, p+1}}-J^{2}_{A_{p-1,p}}}+a_{n_{1}-2}\paren{J^{2}_{A_{p}}J^{2}_{A_{p+1}}-J^{2}_{A_{p-1}}J^{2}_{A_{p}}}}g_{n-n_{1}-\tilde{n}}\overline{S}_{\tilde{n}}
\nonumber\\
+~&\sum_{\substack{p=2\\m_{p}=0}}^{\knm}\bck{\paren{J^{2}_{A_{p-1}}-J^{2}_{A_{p+1}}}\sum_{\tilde{n}=0}^{n-1}h_{n-1-\tilde{n}}\overline{S}_{\tilde{n}}+\paren{J^{2}_{A_{p}}J^{2}_{A_{p+1}}-J^{2}_{A_{p-1}}J^{2}_{A_{p}}}\sum_{\tilde{n}=0}^{n-2}h_{n-2-\tilde{n}}\overline{S}_{\tilde{n}}}
\nonumber\\
-~&J^{2}_{A_{\knm,\knm+1}}\sum_{\tilde{n}=0}^{n-1}h_{n-1-\tilde{n}}\overline{S}_{\tilde{n}}+J^{2}_{A_{\knm+1}}J^{2}_{A_{\knm,\knm+1}}\sum_{\tilde{n}=0}^{n-2}h_{n-2-\tilde{n}}\overline{S}_{\tilde{n}}
\nonumber\\
+~&\paren{J^{2}_{A_{\knm+1}}-J^{2}_{A_{1}}}\sum_{\tilde{n}=0}^{n-1}h_{n-1-\tilde{n}}\overline{S}_{\tilde{n}}+\paren{J^{2}_{A_{\knm}}J^{2}_{A_{\knm,\knm+1}}-J^{2}_{A_{2}}J^{2}_{A_{1,2}}}\sum_{\tilde{n}=0}^{n-2}h_{n-2-\tilde{n}}\overline{S}_{\tilde{n}}
\nonumber\\
=~&\sum_{\tilde{n}=0}^{n}h_{n-\tilde{n}}\overline{S}_{\tilde{n}}-J^{2}_{A_{\knm+1}}\sum_{\tilde{n}=0}^{n-1}h_{n-1-\tilde{n}}\overline{S}_{\tilde{n}}
-\sum_{\tilde{n}=0}^{n}h_{n-\tilde{n}}\overline{S}_{\tilde{n}}
+\sum_{\tilde{n}=0}^{n-1}\paren{J^{2}_{A_{1}}+J^{2}_{A_{1,2}}}h_{n-\tilde{n}-1}\overline{S}_{\tilde{n}}
-\sum_{\tilde{n}=0}^{n-2}J^{2}_{A_{1}}J^{2}_{A_{1,2}}h_{n-\tilde{n}-2}\overline{S}_{\tilde{n}}
\nonumber\\
+~&\sum_{\substack{p=2\\m_{p}\geq 1}}^{\knm}\bck{\paren{J^{2}_{A_{p, p+1}}-J^{2}_{A_{p-1, p}}}\sum_{\tilde{n}=0}^{n-1}h_{n-\tilde{n}-1}\overline{S}_{\tilde{n}}+\paren{J^{2}_{A_{p}}J^{2}_{A_{p+1}}-J^{2}_{A_{p-1}}J^{2}_{A_{p}}}\sum_{\tilde{n}=0}^{n-2}h_{n-\tilde{n}-2}\overline{S}_{\tilde{n}}}
\nonumber\\
+~&\sum_{\substack{p=2\\m_{p}=0}}^{\knm}\bck{\paren{J^{2}_{A_{p-1}}-J^{2}_{A_{p+1}}}\sum_{\tilde{n}=0}^{n-1}h_{n-1-\tilde{n}}\overline{S}_{\tilde{n}}+\paren{J^{2}_{A_{p}}J^{2}_{A_{p+1}}-J^{2}_{A_{p-1}}J^{2}_{A_{p}}}\sum_{\tilde{n}=0}^{n-2}h_{n-2-\tilde{n}}\overline{S}_{\tilde{n}}}
\nonumber\\
-~&J^{2}_{A_{\knm,\knm+1}}\sum_{\tilde{n}=0}^{n-1}h_{n-1-\tilde{n}}\overline{S}_{\tilde{n}}+J^{2}_{A_{\knm+1}}J^{2}_{A_{\knm,\knm+1}}\sum_{\tilde{n}=0}^{n-2}h_{n-2-\tilde{n}}\overline{S}_{\tilde{n}}
\nonumber\\
+~&\paren{J^{2}_{A_{\knm+1}}-J^{2}_{A_{1}}}\sum_{\tilde{n}=0}^{n-1}h_{n-1-\tilde{n}}\overline{S}_{\tilde{n}}+\paren{J^{2}_{A_{\knm}}J^{2}_{A_{\knm,\knm+1}}-J^{2}_{A_{2}}J^{2}_{A_{1,2}}}\sum_{\tilde{n}=0}^{n-2}h_{n-2-\tilde{n}}\overline{S}_{\tilde{n}}
\nonumber
\end{align}
\begin{align}
=~&-J^{2}_{A_{\knm+1}}\sum_{\tilde{n}=0}^{n-1}h_{n-1-\tilde{n}}\overline{S}_{\tilde{n}}
+\paren{J^{2}_{A_{1}}+J^{2}_{A_{1,2}}}\sum_{\tilde{n}=0}^{n-1}h_{n-\tilde{n}-1}\overline{S}_{\tilde{n}}
-J^{2}_{A_{1}}J^{2}_{A_{1,2}}\sum_{\tilde{n}=0}^{n-2}h_{n-\tilde{n}-2}\overline{S}_{\tilde{n}}
\nonumber\\
+~&\sum_{p=2}^{\knm}\bck{\paren{J^{2}_{A_{p-1}}-J^{2}_{A_{p+1}}}\sum_{\tilde{n}=0}^{n-1}h_{n-1-\tilde{n}}\overline{S}_{\tilde{n}}+\paren{J^{2}_{A_{p}}J^{2}_{A_{p+1}}-J^{2}_{A_{p-1}}J^{2}_{A_{p}}}\sum_{\tilde{n}=0}^{n-2}h_{n-2-\tilde{n}}\overline{S}_{\tilde{n}}}
\nonumber\\
-~&J^{2}_{A_{\knm,\knm+1}}\sum_{\tilde{n}=0}^{n-1}h_{n-1-\tilde{n}}\overline{S}_{\tilde{n}}+J^{2}_{A_{\knm+1}}J^{2}_{A_{\knm,\knm+1}}\sum_{\tilde{n}=0}^{n-2}h_{n-2-\tilde{n}}\overline{S}_{\tilde{n}}
\nonumber\\
+~&\paren{J^{2}_{A_{\knm+1}}-J^{2}_{A_{1}}}\sum_{\tilde{n}=0}^{n-1}h_{n-1-\tilde{n}}\overline{S}_{\tilde{n}}+\paren{J^{2}_{A_{\knm}}J^{2}_{A_{\knm,\knm+1}}-J^{2}_{A_{2}}J^{2}_{A_{1,2}}}\sum_{\tilde{n}=0}^{n-2}h_{n-2-\tilde{n}}\overline{S}_{\tilde{n}}
\nonumber\\
=~&-J^{2}_{A_{\knm+1}}\sum_{\tilde{n}=0}^{n-1}h_{n-1-\tilde{n}}\overline{S}_{\tilde{n}}
+\paren{J^{2}_{A_{1}}+J^{2}_{A_{1,2}}}\sum_{\tilde{n}=0}^{n-1}h_{n-\tilde{n}-1}\overline{S}_{\tilde{n}}
-J^{2}_{A_{1}}J^{2}_{A_{1,2}}\sum_{\tilde{n}=0}^{n-2}h_{n-\tilde{n}-2}\overline{S}_{\tilde{n}}
\nonumber\\
+~&\paren{J^{2}_{A_{1}}+J^{2}_{A_{2}}-J^{2}_{A_{\knm}}-J^{2}_{A_{\knm+1}}}\sum_{\tilde{n}=0}^{n-1}h_{n-1-\tilde{n}}\overline{S}_{\tilde{n}}+\paren{J^{2}_{A_{\knm}}J^{2}_{A_{\knm+1}}-J^{2}_{A_{1}}J^{2}_{A_{2}}}\sum_{\tilde{n}=0}^{n-2}h_{n-2-\tilde{n}}\overline{S}_{\tilde{n}}
\nonumber\\
-~&J^{2}_{A_{\knm,\knm+1}}\sum_{\tilde{n}=0}^{n-1}h_{n-1-\tilde{n}}\overline{S}_{\tilde{n}}+J^{2}_{A_{\knm+1}}J^{2}_{A_{\knm,\knm+1}}\sum_{\tilde{n}=0}^{n-2}h_{n-2-\tilde{n}}\overline{S}_{\tilde{n}}
\nonumber\\
+~&\paren{J^{2}_{A_{\knm+1}}-J^{2}_{A_{1}}}\sum_{\tilde{n}=0}^{n-1}h_{n-1-\tilde{n}}\overline{S}_{\tilde{n}}+\paren{J^{2}_{A_{\knm}}J^{2}_{A_{\knm,\knm+1}}-J^{2}_{A_{2}}J^{2}_{A_{1,2}}}\sum_{\tilde{n}=0}^{n-2}h_{n-2-\tilde{n}}\overline{S}_{\tilde{n}}
\nonumber\\
=~&\paren{J^{2}_{X}+J^{2}_{Y}+J^{2}_{Z}-J^{2}_{X}-J^{2}_{Y}-J^{2}_{Z}}\sum_{\tilde{n}=0}^{n-1}h_{n-1-\tilde{n}}\overline{S}_{\tilde{n}}
+\paren{J^{2}_{X}J^{2}_{Y}+J^{2}_{Y}J^{2}_{Z}
+J^{2}_{Z}J^{2}_{X}-J^{2}_{X}J^{2}_{Y}-J^{2}_{Y}J^{2}_{Z}-J^{2}_{Z}J^{2}_{X}}\sum_{\tilde{n}=0}^{n-2}h_{n-2-\tilde{n}}\overline{S}_{\tilde{n}}
\nonumber\\
=~&0,
\end{align}
therefore, Eq.~\eqref{supeq:conditionmhole1} is satisfied. Subtracting the left-hand side of Eq.~\eqref{supeq:conditionmhole1} from that of Eq.~\eqref{supeq:conditionmhole2}, we obtain
\begin{align}
-~&\sum_{\tilde{n}=0}^{n}h_{n-\tilde{n}}\overline{S}_{\tilde{n}}+J^{2}_{A_{1}}\sum_{\tilde{n}=0}^{n-1}h_{n-1-\tilde{n}}\overline{S}_{\tilde{n}}
+J^{2}_{A_{1,2}}\sum_{\tilde{n}=0}^{n-1}h_{n-1-\tilde{n}}\overline{S}_{\tilde{n}}
-J^{2}_{A_{1}}J^{2}_{A_{1,2}}\sum_{\tilde{n}=0}^{n-2}h_{n-2-\tilde{n}}\overline{S}_{\tilde{n}}
+\sum_{\tilde{n}=0}^{n}\sum_{n_{1}=0}^{\tilde{n}}J^{2n_{1}}_{A_{2}}g_{n-\tilde{n}}\overline{S}_{\tilde{n}-n_{1}}
\nonumber\\
=-~&\sum_{\tilde{n}=0}^{n}h_{n-\tilde{n}}\overline{S}_{\tilde{n}}+J^{2}_{A_{1}}\sum_{\tilde{n}=0}^{n-1}h_{n-1-\tilde{n}}\overline{S}_{\tilde{n}}
+J^{2}_{A_{1,2}}\sum_{\tilde{n}=0}^{n-1}h_{n-1-\tilde{n}}\overline{S}_{\tilde{n}}
-J^{2}_{A_{1}}J^{2}_{A_{1,2}}\sum_{\tilde{n}=0}^{n-1}h_{n-1-\tilde{n}}\overline{S}_{\tilde{n}}
\nonumber\\
+~&\sum_{\tilde{n}=0}^{n}h_{n-\tilde{n}}\overline{S}_{\tilde{n}}
-\sum_{\tilde{n}=0}^{n-1}\paren{J^{2}_{A_{1}}+J^{2}_{A_{1,2}}}h_{n-\tilde{n}-1}\overline{S}_{\tilde{n}}
+\sum_{\tilde{n}=0}^{n-2}J^{2}_{A_{1}}J^{2}_{A_{1,2}}h_{n-\tilde{n}-2}\overline{S}_{\tilde{n}}=0,
\end{align}
therefore, Eq.~\eqref{supeq:conditionmhole2} is satisfied. Subtracting the left-hand side of Eq.~\eqref{supeq:conditionmhole3} from that of Eq.~\eqref{supeq:conditionmhole2}, we obtain
\begin{align}
&\sum_{\tilde{n}=0}^{n}h_{n-\tilde{n}}\overline{S}_{\tilde{n}}
-J^{2}_{A_{\knm+1}}\sum_{\tilde{n}=0}^{n-1}h_{n-1-\tilde{n}}\overline{S}_{\tilde{n}}
-J^{2}_{A_{\knm,\knm+1}}\sum_{\tilde{n}=0}^{n-1}h_{n-1-\tilde{n}}\overline{S}_{\tilde{n}}
\nonumber\\
+~&J^{2}_{A_{\knm+1}}J^{2}_{A_{\knm,\knm+1}}\sum_{\tilde{n}=0}^{n-2}h_{n-2-\tilde{n}}\overline{S}_{\tilde{n}}
-\sum_{\tilde{n}=0}^{n}\sum_{n_{1}=0}^{\tilde{n}}J^{2n_{1}}_{A_{\knm}}g_{n-\tilde{n}}\overline{S}_{\tilde{n}-n_{1}}
\nonumber\\
=~&\sum_{\tilde{n}=0}^{n}h_{n-\tilde{n}}\overline{S}_{\tilde{n}}
-J^{2}_{A_{\knm+1}}\sum_{\tilde{n}=0}^{n-1}h_{n-1-\tilde{n}}\overline{S}_{\tilde{n}}
-J^{2}_{A_{\knm,\knm+1}}\sum_{\tilde{n}=0}^{n-1}h_{n-1-\tilde{n}}\overline{S}_{\tilde{n}}
\nonumber\\
+~&J^{2}_{A_{\knm+1}}J^{2}_{A_{\knm,\knm+1}}\sum_{\tilde{n}=0}^{n-2}h_{n-2-\tilde{n}}\overline{S}_{\tilde{n}}
-\sum_{\tilde{n}=0}^{n}h_{n-\tilde{n}}\overline{S}_{\tilde{n}}
\nonumber\\
+~&\sum_{\tilde{n}=0}^{n-1}\paren{J^{2}_{A_{\knm+1}}+J^{2}_{A_{\knm,\knm+1}}}h_{n-\tilde{n}-1}\overline{S}_{\tilde{n}}
-\sum_{\tilde{n}=0}^{n-2}J^{2}_{A_{\knm+1}}J^{2}_{A_{\knm,\knm+1}}h_{n-\tilde{n}-2}\overline{S}_{\tilde{n}}=0,
\end{align}
therefore, Eq.~\eqref{supeq:conditionmhole3} is satisfied.

We next prove that Eq.~\eqref{supeq:conditionlrhole} is satisfied.
$R^{k-2(n-1)-(m+1),m+1}$ and $R^{k-2(n-1)-(m+2),m+2}$ for $m\geq 1$ can be expressed as 
\begin{align}
R^{k-2(n-1)-(m+1),m+1}&=g_{0}S_{n-1}+g_{1}S_{n-2}+\cdots+g_{n-3}S_{2}+g_{n-2}S_{1}+g_{n-1}S_{0},
\label{supeq:k-2(n-1)-(m+1)m+1lrhole}\\
R^{k-2(n-1)-(m+2),m+2}&=h_{0}S_{n-1}+h_{1}S_{n-2}+\cdots+h_{n-3}S_{2}+h_{n-2}S_{1}+h_{n-1}S_{0},
\label{supeq:k-2(n-1)-(m+2)m+2lrhole}
\end{align}
where $g_{0}=h_{0}=1$ and 
$h_{l}=\sum_{\tilde{l}=0}^{l}a_{l-\tilde{l}}g_{\tilde{l}}$
are satisfied.
$R^{k-2n-m, m}$ is expressed as
\begin{align}
R^{k-2n-m,m}&=g_{0}S_{n}+g_{1}S_{n-1}+\cdots+g_{n-3}S_{3}+g_{n-2}S_{2}+g_{n-1}S_{1}+g_{n}.
\label{supeq:k-2n-mmlrhole}
\end{align}
In this proof, we use a notation $\overline{S}_{p}\equiv S_{p}\paren{A_{1}A_{2}\cdots A_{k-2n-2m-2}}$.
In a similar manner to the case of Eq.~\eqref{supeq:conditionmhole1}, the left-hand side of Eq.~\eqref{supeq:conditionlrhole} becomes
\begin{align}
&\sum_{\tilde{n}=0}^{n}\sum_{n_{1}=0}^{\tilde{n}}\paren{J^{2n_{1}}_{A_{\knm-2}}-J^{2n_{1}}_{A_{2}}}g_{n-\tilde{n}}\overline{S}_{\tilde{n}-n_{1}}
\nonumber\\
+~&\sum_{p=2}^{\knm-2}\bck{\paren{J^{2}_{A_{p-1}}-J^{2}_{A_{p+1}}}\sum_{\tilde{n}=0}^{n-1}h_{n-1-\tilde{n}}\overline{S}_{\tilde{n}}+\paren{J^{2}_{A_{p}}J^{2}_{A_{p+1}}-J^{2}_{A_{p-1}}J^{2}_{A_{p}}}\sum_{\tilde{n}=0}^{n-2}h_{n-2-\tilde{n}}\overline{S}_{\tilde{n}}}
\nonumber\\
+~&\sum_{\tilde{n}=0}^{n-1}\sum_{n_{1}=0}^{\tilde{n}}\paren{J^{2(n_{1}+1)}_{A_{2}}+J^{2(n_{1}+1)}_{A_{1,2}}-J^{2(n_{1}+1)}_{A_{\knm-2}}-J^{2(n_{1}+1)}_{A_{\knm-2,\knm-1}}}g_{n-1-\tilde{n}}\overline{S}_{\tilde{n}-n_{1}}
\nonumber\\
=~&\sum_{\tilde{n}=0}^{n-1}\paren{J^{2}_{A_{1}}+J^{2}_{A_{1,2}}-J^{2}_{A_{\knm-1}}-J^{2}_{A_{\knm-2,\knm-1}}}h_{n-1-\tilde{n}}\overline{S}_{\tilde{n}}
\nonumber\\
+~&\sum_{\tilde{n}=0}^{n-2}\paren{J^{2}_{A_{\knm-1}}J^{2}_{A_{\knm-2,\knm-1}}-J^{2}_{A_{1}}J^{2}_{A_{1,2}}}h_{n-2-\tilde{n}}\overline{S}_{\tilde{n}}
\nonumber\\
+~&\paren{J^{2}_{A_{1}}+J^{2}_{A_{2}}-J^{2}_{A_{\knm-2}}-J^{2}_{A_{\knm-1}}}\sum_{\tilde{n}=0}^{n-1}h_{n-1-\tilde{n}}\overline{S}_{\tilde{n}}+\paren{J^{2}_{A_{\knm-2}}J^{2}_{A_{\knm-1}}-J^{2}_{A_{1}}J^{2}_{A_{2}}}\sum_{\tilde{n}=0}^{n-2}h_{n-2-\tilde{n}}\overline{S}_{\tilde{n}}
\nonumber\\
+~&\paren{J^{2}_{A_{\knm-1}}-J^{2}_{A_{1}}}\sum_{\tilde{n}=0}^{n-1}h_{n-1-\tilde{n}}\overline{S}_{\tilde{n}}+\paren{J^{2}_{A_{\knm-2}}J^{2}_{A_{\knm-2,\knm-1}}-J^{2}_{A_{2}}J^{2}_{A_{1,2}}}\sum_{\tilde{n}=0}^{n-2}h_{n-2-\tilde{n}}\overline{S}_{\tilde{n}}
\nonumber\\
=~&\paren{J^{2}_{X}+J^{2}_{Y}+J^{2}_{Z}-J^{2}_{X}-J^{2}_{Y}-J^{2}_{Z}}\sum_{\tilde{n}=0}^{n-1}h_{n-1-\tilde{n}}\overline{S}_{\tilde{n}}
\nonumber\\
+~&\paren{J^{2}_{X}J^{2}_{Y}+J^{2}_{Y}J^{2}_{Z}
+J^{2}_{Z}J^{2}_{X}-J^{2}_{X}J^{2}_{Y}-J^{2}_{Y}J^{2}_{Z}-J^{2}_{Z}J^{2}_{X}}\sum_{\tilde{n}=0}^{n-2}h_{n-2-\tilde{n}}\overline{S}_{\tilde{n}}=0,
\end{align}
therefore, Eq.~\eqref{supeq:conditionlrhole} is satisfied. As a result, all the conditions Eqs.~\eqref{supeq:condition}-\eqref{supeq:conditionlrhole} are satisfied.

\subsection{E. Coefficients of $Q_{k}$ in closed form}
In this subsection, we derive the coefficients of $Q_{k}$ in closed form, namely, 
Eqs.~(14)-(20).
As shown in the previous subsection, the function $r$ for $(k-2n-m,m)$ operators can be written as
\begin{align}
r^{k-2n-m,m}\paren{\overline{A_{1}^{1+m_{1}}A_{2}^{1+m_{2}}\cdots A_{k-2n-2m-1}^{1+m_{k-2n-2m-1}}}}
&\equiv R^{k-2n-m,m}\paren{A_{1}A_{2}\cdots A_{k-2n-2m-1}}
\nonumber\\
&=\sum_{\tilde{n}=0}^{n}g^{k-2n-m,m}_{n-\tilde{n}}S_{\tilde{n}}\paren{A_{1}A_{2}\cdots A_{k-2n-2m-1}}.
\end{align}
We define $f\paren{n, m}\equiv g^{k-2n-m, m}_{n}$. Then, from the recursive way in the previous subsection,
\begin{align}
R^{k-2n-m,m}\paren{A_{1}A_{2}\cdots A_{k-2n-2m-1}}
=\sum_{\tilde{n}=0}^{n}f\paren{n-\tilde{n},m+\tilde{n}}S_{\tilde{n}}\paren{A_{1}A_{2}\cdots A_{k-2n-2m-1}},
\end{align}
where the function $f$ is determined as
\begin{align}
f\paren{0,m}&=1,
\label{supeq:f1}\\
f\paren{n,0}&=0\quad (n\geq 1),
\label{supeq:f2}\\
f\paren{n,m}&=\sum_{p=1}^{n}\sum_{\tilde{m}=1}^{m}a_{p}f\paren{n-p,p+\tilde{m}-1} \quad (n\geq 1~\text{and}~m\geq 1).
\label{supeq:f3}
\end{align}
We prove that 
Eq.~(20)
satisfies Eqs.~\eqref{supeq:f1}-\eqref{supeq:f3}. Obviously, Eqs.~\eqref{supeq:f1}-\eqref{supeq:f2} are satisfied. Substituting
Eq.~(20)
into the right hand side of Eq.~\eqref{supeq:f3}, we obtain
\begin{align}
\sum_{p=1}^{n}\sum_{\tilde{m}=1}^{m}a_{p}\frac{p+\tilde{m}-1}{n+\tilde{m}-1}\sum_{p_{1}=1}^{n-p}\binom{n+\tilde{m}-1}{p_{1}}\sum_{\substack{j1,j2,\ldots,jp_{1}\geq 1\\\\j1+j2+\cdots+jp_{1}=n-p}}a_{j1}a_{j2}\cdots a_{jp_{1}}.
\end{align}
Let us focus on the coefficient of $a^{n_{1}}_{j_{1}}a^{n_{2}}_{j_{2}}\cdots a^{n_{c}}_{j_{c}}$, where $j_{1}<j_{2}<\cdots <j_{c}$, $n_{1}+n_{2}+\cdots +n_{c}=N$, and $n_{1}j_{1}+n_{2}j_{2}+\cdots + n_{c}j_{c}=n$. It is given as
\begin{align}
&\sum_{\tilde{m}=1}^{m}\sum_{l=1}^{c}\frac{j_{l}+\tilde{m}-1}{n+\tilde{m}-1}
\binom{n+\tilde{m}-1}{N-1}\frac{n_{l}(N-1)!}{n_{1}!n_{2}!\cdots n_{c}!}
=\sum_{\tilde{m}=1}^{m}\frac{n+\paren{\tilde{m}-1}N}{n+\tilde{m}-1}\cdot \frac{(n+\tilde{m}-1)!}{(n+\tilde{m}-N)! n_{1}!n_{2}!\cdots n_{c}!}
\nonumber\\
=~&\frac{m(n+m-1)!}{(n+m-N)!n_{1}!n_{2}!\cdots n_{c}!}.
\end{align}
On the other hand, the coefficient of $a^{n_{1}}_{j_{1}}a^{n_{2}}_{j_{2}}\cdots a^{n_{c}}_{j_{c}}$ in $f\paren{n, m}$ is 
\begin{align}
\frac{m}{n+m}\binom{n+m}{N}\frac{N!}{n_{1}!n_{2}!\cdots n_{c}!}=\frac{m(n+m-1)!}{(n+m-N)!n_{1}!n_{2}!\cdots n_{c}!},
\end{align}
therefore, 
Eq.~(20)
satisfies Eq.~\eqref{supeq:f3}.

\section*{S2. Explicit proof of $[Q_{k},Z]=0$ in the case of the XXZ chain}
For the case of the $XXZ$ chain ($J_{X}=J_{Y}$), we prove explicitly that $Q_{k}$ for $k\geq 2$ is commutative with a magnetic field in the z-axis direction, \textit{i.e.}, 
$[Q_{k},Z]=0$. To prove this, we focus on $(l,m)$ operators in $Q_{k}$. First, $\overline{ZZ\cdots Z}=\overline{Z^{l-m-1}}$ is commutative with $Z$ obviously.
For the other $(l,m)$ operators $\overline{A_{1}^{1+m_{1}}A_{2}^{1+m_{2}}\cdots A_{l-m-1}^{1+m_{l-m-1}}}$, at least one character in $A_{1}$, $A_{2}$, \ldots, $A_{l-m-1}$ is $X$ or $Y$, therefore, each $(l,m)$ operator can be expressed as at least one of the following forms:
\begin{align}
&\overline{C_{1}^{1+m_{1}}C_{2}^{1+m_{2}}\cdots C_{\alpha}^{1+m_{\alpha}}Z^{1+m_{\alpha+1}}\cdots},
\label{supeq:cz}\\
&\overline{\cdots Z^{1+m_{\alpha-1}}C_{\alpha}^{1+m_{\alpha}}C_{\alpha+1}^{1+m_{\alpha+1}}\cdots C_{l-m-1}^{1+m_{l-m-1}}},
\label{supeq:zc}\\
&\overline{\cdots Z^{1+m_{\alpha-1}}C_{\alpha}^{1+m_{\alpha}}C_{\alpha+1}^{1+m_{\alpha+1}}\cdots C_{\beta}^{1+m_{\beta}}Z^{1+m_{\beta+1}}\cdots},
\label{supeq:zcz}
\end{align}
where $C_{j}\in \bce{X,Y}$ for all $j$. In addition, we define $D_{j}$ as $\bce{C_{j},D_{j}}=\bce{X, Y}$.
In the case of Eq.~\eqref{supeq:cz}, we consider a commutator
\begin{align}
\begin{array}{cccc}
\cline{1-4}
C_{1}^{1+m_{1}}C_{2}^{1+m_{2}}&\cdots&C_{\alpha}^{1+m_{\alpha}}~Z^{1+m_{\alpha+1}}&\cdots\\
&&\!\!\!\!\!\!\!Z&
\end{array}
&=
\begin{array}{ccc}
C_{1}I^{m_{1}}Z I^{m_{2}}\cdots I^{m_{\alpha}}&D_{\alpha}&I^{m_{\alpha+1}}\cdots\\
&Z&
\end{array}
\nonumber\\
&=s\paren{D_{\alpha}Z}
\begin{array}{c}
C_{1}I^{m_{1}}Z I^{m_{2}}\cdots I^{m_{\alpha}}C_{\alpha}I^{m_{\alpha+1}}\cdots
\end{array}.
\end{align}
The same operator is also generated from
\begin{align}
\begin{array}{cccccc}
\cline{2-6}
&D_{1}^{1+m_{1}}&D_{2}^{1+m_{2}}&\cdots&D_{\alpha}^{1+m_{\alpha}}~Z^{1+m_{\alpha+1}}&\cdots\\
&\hspace{-1.2cm}Z&&&&
\end{array}
&=
\begin{array}{ccc}
D_{1}&I^{m_{1}}Z I^{m_{2}}\cdots I^{m_{\alpha}}C_{\alpha}I^{m_{\alpha+1}}\cdots\\
\!\!Z&
\end{array}
\nonumber\\
&=s\paren{D_{1}Z}
\begin{array}{c}
C_{1}I^{m_{1}}Z I^{m_{2}}\cdots I^{m_{\alpha}}C_{\alpha}I^{m_{\alpha+1}}\cdots
\end{array}.
\end{align}
We prove that these two terms in $[Q_{k},Z]$ are cancelled.
In the case of $J_{X}=J_{Y}$, $r^{l, m}\paren{\overline{C_{1}^{1+m_{1}}C_{2}^{1+m_{2}}\cdots C_{\alpha}^{1+m_{\alpha}}Z^{1+m_{\alpha}}\cdots}}=r^{l, m}\paren{\overline{D_{1}^{1+m_{1}}D_{2}^{1+m_{2}}\cdots D_{\alpha}^{1+m_{\alpha}}Z^{1+m_{\alpha}}\cdots}}$, and therefore, 
\begin{align}
&s\paren{D_{\alpha}Z}q^{l, m}_{\overline{C_{1}^{1+m_{1}}C_{2}^{1+m_{2}}\cdots C_{\alpha}^{1+m_{\alpha}}Z^{1+m_{\alpha}}\cdots}}+
s\paren{D_{1}Z}q^{l, m}_{\overline{D_{1}^{1+m_{1}}D_{2}^{1+m_{2}}\cdots D_{\alpha}^{1+m_{\alpha}}Z^{1+m_{\alpha}}\cdots}}
\nonumber\\
\propto~&s\paren{D_{\alpha}Z}s\paren{C_{1}C_{2}\cdots C_{\alpha}Z}+s\paren{D_{1}Z}s\paren{D_{1}D_{2}\cdots D_{\alpha}Z}
\nonumber\\
=~&-s\paren{C_{1}C_{2}\cdots C_{\alpha}}+s\paren{D_{1}Z}s\paren{D_{\alpha}Z}s\paren{D_{1}D_{2}\cdots D_{\alpha}}.
\end{align}
By using $s\paren{XY}=-s\paren{YX}=1$ and $s\paren{XYX}=s\paren{YXY}=-1$,
\begin{align}
s\paren{C_{1}C_{2}\cdots C_{\alpha}}/s\paren{D_{1}D_{2}\cdots D_{\alpha}}=
\begin{cases}
+1& (D_{1}=D_{\alpha}),\\
-1& (D_{1}\neq D_{\alpha}),
\end{cases}
\label{supeq:ddpm1}
\end{align}
and in both cases,
\begin{align}
s\paren{D_{1}Z}s\paren{D_{\alpha}Z}s\paren{D_{1}D_{2}\cdots D_{\alpha}}=s\paren{C_{1}C_{2}\cdots C_{\alpha}}.
\end{align}
Therefore, 
\begin{align}
s\paren{D_{\alpha}Z}q^{l, m}_{\overline{C_{1}^{1+m_{1}}C_{2}^{1+m_{2}}\cdots C_{\alpha}^{1+m_{\alpha}}Z^{1+m_{\alpha}}\cdots}}+
s\paren{D_{1}Z}q^{l, m}_{\overline{D_{1}^{1+m_{1}}D_{2}^{1+m_{2}}\cdots D_{\alpha}^{1+m_{\alpha}}Z^{1+m_{\alpha}}\cdots}}=0
\end{align}
is satisfied.

In the case of Eq.~\eqref{supeq:zc}, we consider two commutators
\begin{align}
\begin{array}{ccccc}
\cline{1-5}
\cdots&Z^{1+m_{\alpha-1}}~C_{\alpha}^{1+m_{\alpha}}&C_{\alpha+1}^{1+m_{\alpha+1}}&\cdots&C_{l-m-1}^{1+m_{l-m-1}}\\
&\!\!\!\!\!\!\!Z&&&
\end{array}
&=
\begin{array}{ccc}
\cdots I^{m_{\alpha-1}}&D_{\alpha}&I^{m_{\alpha}}Z I^{m_{\alpha+1}}\cdots I^{m_{l-m-1}}C_{l-m-1}\\
&Z&
\end{array}
\nonumber\\
&=s\paren{D_{\alpha}Z}
\begin{array}{c}
\cdots I^{m_{\alpha-1}}C_{\alpha}I^{m_{\alpha}}Z I^{m_{\alpha+1}}\cdots I^{m_{l-m-1}}C_{l-m-1}
\end{array},
\end{align}
and
\begin{align}
\begin{array}{ccccc}
\cline{1-5}
\cdots&Z^{1+m_{\alpha-1}}~D_{\alpha}^{1+m_{\alpha}}&D_{\alpha+1}^{1+m_{\alpha+1}}&\cdots&D_{l-m-1}^{1+m_{l-m-1}}\\
&&&&\hspace{1.5cm}Z
\end{array}
&=
\begin{array}{cc}
\cdots I^{m_{\alpha-1}}C_{\alpha}I^{m_{\alpha}}Z I^{m_{\alpha+1}}\cdots I^{m_{l-m-1}}&D_{l-m-1}\\
&\!\!\!\!\!\!\!\!\!\!\!\!\!\!\!\!Z
\end{array}
\nonumber\\
&=s\paren{D_{l-m-1}Z}
\begin{array}{c}
\cdots I^{m_{\alpha-1}}C_{\alpha}I^{m_{\alpha}}Z I^{m_{\alpha+1}}\cdots I^{m_{l-m-1}}C_{l-m-1}
\end{array}.
\end{align}
In a similar manner to the case of Eq.~\eqref{supeq:cz}, 
\begin{align}
s\paren{D_{\alpha}Z}q^{l, m}_{\overline{\cdots Z^{1+m_{\alpha-1}}C_{\alpha}^{1+m_{\alpha}}C_{\alpha+1}^{1+m_{\alpha+1}}\cdots C_{l-m-1}^{1+m_{l-m-1}}}}+
s\paren{D_{l-m-1}Z}q^{l, m}_{\overline{\cdots Z^{1+m_{\alpha-1}}D_{\alpha}^{1+m_{\alpha}}D_{\alpha+1}^{1+m_{\alpha+1}}\cdots D_{l-m-1}^{1+m_{l-m-1}}}}=0,
\end{align}
where we have used
\begin{align}
s\paren{D_{\alpha}Z}s\paren{ZC_{\alpha}C_{\alpha+1}\cdots C_{l-m-1}}=s\paren{D_{l-m-1}Z}s\paren{ZD_{\alpha+1}D_{\alpha}\cdots D_{l-m-1}}.
\end{align}

In the case of Eq.~\eqref{supeq:zcz}, we consider four commutators
\begin{align}
\begin{array}{ccccc}
\cline{1-5}
\cdots&Z^{1+m_{\alpha-1}}~C_{\alpha}^{1+m_{\alpha}}&C_{\alpha+1}^{1+m_{\alpha+1}}&\cdots&C_{\beta}^{1+m_{\beta}}~Z^{1+m_{\beta+1}}\cdots\\
&\!\!\!\!\!\!\!Z&&&
\end{array}
&=
\begin{array}{ccc}
\cdots I^{m_{\alpha-1}}&D_{\alpha}&I^{m_{\alpha}}Z I^{m_{\alpha+1}}\cdots I^{m_{\beta}}D_{\beta}I^{m_{\beta+1}}\cdots\\
&Z&
\end{array}
\nonumber\\
&=s\paren{D_{\alpha}Z}
\begin{array}{c}
\cdots I^{m_{\alpha-1}}C_{\alpha}I^{m_{\alpha}}Z I^{m_{\alpha+1}}\cdots I^{m_{\beta}}D_{\beta}I^{m_{\beta+1}}\cdots
\end{array},
\label{supeq:zcz1}\\
\begin{array}{ccccc}
\cline{1-5}
\cdots&Z^{1+m_{\alpha-1}}~D_{\alpha}^{1+m_{\alpha}}&D_{\alpha+1}^{1+m_{\alpha+1}}&\cdots&D_{\beta}^{1+m_{\beta}}~Z^{1+m_{\beta+1}}\cdots\\
&&&&\!\!\!\!\!\!\!\!\!\!\!\!\!\!\!Z
\end{array}
&=
\begin{array}{ccc}
\cdots I^{m_{\alpha-1}}C_{\alpha}I^{m_{\alpha}}Z I^{m_{\alpha+1}}\cdots I^{m_{\beta}}&C_{\beta}& I^{m_{\beta+1}}\cdots\\
&Z&
\end{array}
\nonumber\\
&=s\paren{C_{\beta}Z}
\begin{array}{c}
\cdots I^{m_{\alpha-1}}C_{\alpha}I^{m_{\alpha}}Z I^{m_{\alpha+1}}\cdots I^{m_{\beta}}D_{\beta}I^{m_{\beta+1}}\cdots
\end{array},
\label{supeq:zcz2}\\
\begin{array}{ccccc}
\cline{1-5}
\cdots&Z^{1+m_{\alpha-1}}~D_{\alpha}^{1+m_{\alpha}}&D_{\alpha+1}^{1+m_{\alpha+1}}&\cdots&D_{\beta}^{1+m_{\beta}}~Z^{1+m_{\beta+1}}\cdots\\
&\!\!\!\!\!\!\!Z&&&
\end{array}
&=
\begin{array}{ccc}
\cdots I^{m_{\alpha-1}}&C_{\alpha}&I^{m_{\alpha}}Z I^{m_{\alpha+1}}\cdots I^{m_{\beta}}C_{\beta}I^{m_{\beta+1}}\cdots\\
&Z&
\end{array}
\nonumber\\
&=s\paren{C_{\alpha}Z}
\begin{array}{c}
\cdots I^{m_{\alpha-1}}D_{\alpha}I^{m_{\alpha}}Z I^{m_{\alpha+1}}\cdots I^{m_{\beta}}C_{\beta}I^{m_{\beta+1}}\cdots
\end{array},
\label{supeq:zcz3}\\
\begin{array}{ccccc}
\cline{1-5}
\cdots&Z^{1+m_{\alpha-1}}~C_{\alpha}^{1+m_{\alpha}}&C_{\alpha+1}^{1+m_{\alpha+1}}&\cdots&C_{\beta}^{1+m_{\beta}}~Z^{1+m_{\beta+1}}\cdots\\
&&&&\!\!\!\!\!\!\!\!\!\!\!\!\!\!\!Z
\end{array}
&=
\begin{array}{ccc}
\cdots I^{m_{\alpha-1}}D_{\alpha}I^{m_{\alpha}}Z I^{m_{\alpha+1}}\cdots I^{m_{\beta}}&D_{\beta}& I^{m_{\beta+1}}\cdots\\
&Z&
\end{array}
\nonumber\\
&=s\paren{D_{\beta}Z}
\begin{array}{c}
\cdots I^{m_{\alpha-1}}D_{\alpha}I^{m_{\alpha}}Z I^{m_{\alpha+1}}\cdots I^{m_{\beta}}C_{\beta}I^{m_{\beta+1}}\cdots
\end{array}.
\label{supeq:zcz4}
\end{align}
Two commutators Eqs.\eqref{supeq:zcz1}-\eqref{supeq:zcz2} in $Q_{k}$ are cancelled from the identity
\begin{align}
s\paren{D_{\alpha}Z}s\paren{ZC_{\alpha}C_{\alpha+1}\cdots C_{\beta}Z}
+s\paren{C_{\beta}Z}s\paren{ZD_{\alpha}D_{\alpha+1}\cdots D_{\beta}Z}=0,
\end{align}
which is proved from Eq.~\eqref{supeq:ddpm1}. In a similar manner, Two commutators Eqs.\eqref{supeq:zcz3}-\eqref{supeq:zcz4} in $Q_{k}$ are cancelled. Therefore, all the commutators in $[Q_{k},Z]$ are cancelled, and $[Q_{k},Z]=0$.

\section*{S3. Case of the XXX chain}
\begin{figure}[b]
\centering
\includegraphics[width=0.8\columnwidth,clip]{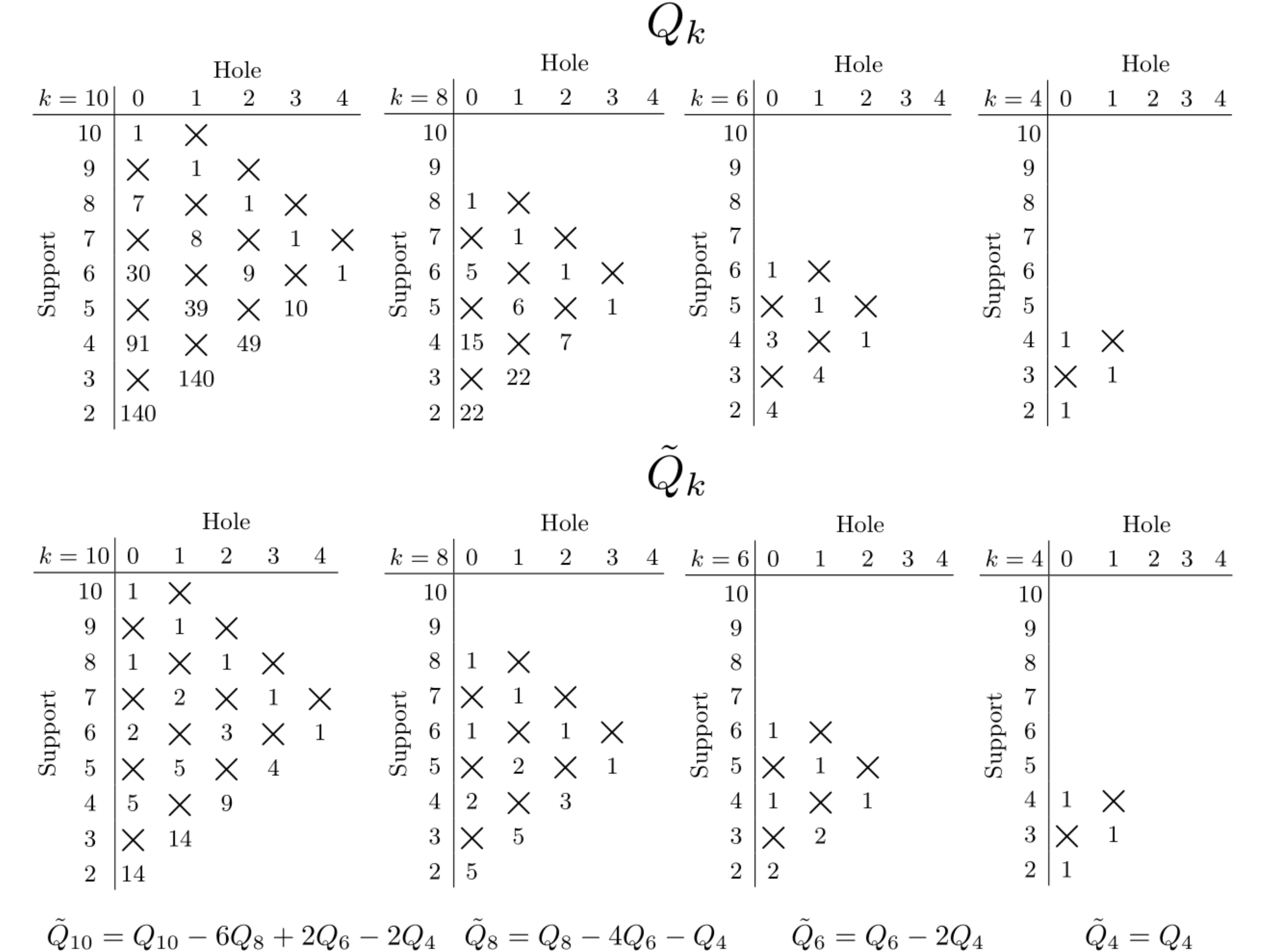}
\caption{Values of $R^{k-2n-m, m}$ for $Q_{k}$ and $\tilde{R}^{k-2n-m, m}$ for $\tilde{Q}_{k}$, respectively in the XXX chain. $\tilde{Q}_{10},~\tilde{Q}_{8},~\tilde{Q}_{6}$, and $\tilde{Q}_{4}$ are linear combinations of $Q_{10},~Q_{8},~Q_{6}$, and $Q_{4}$.  The support is $k-2n-m$, and the hole is $m$.}
\label{supfig:tildeq}
\end{figure}

In this section, we consider the case of the XXX chain. Without loss of generality, we can set $J_{X}=J_{Y}=J_{Z}=1$. In this case, the coefficients of the conserved quantities $Q_{k}$ we obtained in 
Eqs.~(15)-(20) becomes
\begin{gather}
q^{k-2n-m,m}_{\overline{A_{1}^{1+m_{1}}A_{2}^{1+m_{2}}\cdots A_{k-2n-2m-1}^{1+m_{k-2n-2m-1}}}}
= s\paren{A_{1}A_{2}\cdots A_{k-2n-2m-1}}R^{k-2n-m, m},
\\
R^{k-2n-m,m}
=\sum_{\tilde{n}=0}^{n}f\paren{n-\tilde{n},m+\tilde{n}}S_{\tilde{n}},
\label{supeq:defxxxR}\\
f\paren{0,m}=1, \quad
f\paren{n,m}=
\frac{m}{n+m}\sum_{p=1}^{n}\binom{n+m}{p}\sum_{\substack{j1,j2,\ldots,jp\geq 1\\\\j1+j2+\cdots+jp=n}}\binom{j1+2}{2}\binom{j2+2}{2}\cdots \binom{jp+2}{2} \quad (n\geq 1),
\\
S_{0}= 1, \quad
S_{p}= \sum_{1\leq j1\leq j2\leq\cdots\leq jp\leq l} 1 \quad (p\geq 1),
\end{gather}
where we have used $a_{n}=\binom{n+2}{2}$.
One main difference is that $R^{k-2n-m,m}$ does not depend on $A_{1}A_{2}\cdots A_{k-2n-2m-1}$. Using this property, we prove that 
\begin{align}
R^{k-2n-m,m}=R^{k-2n-(m-1),m-1}+R^{k-2(n-1)-(m+1),m+1} \quad \text{for}~m\geq 1,
\label{supeq:xxxR}
\end{align}
which is discussed in the main text. To prove it, instead of calculating Eq.~\eqref{supeq:defxxxR} explicitly, we consider Eq.~\eqref{supeq:conditionmhole1}, which is one of the conditions $R^{k-2n-m,m}$ satisfies.
Substituting $J_{X}=J_{Y}=J_{Z}=1$ into 
Eq.~\eqref{supeq:conditionmhole1}
and using the property of $R^{k-2n-m,m}$, many terms are cancelled and we obtain
\begin{align}
&R^{k-2n-m,m}-R^{k-2n-m+1,m-1}-J^{2}_{A_{k-2n-2m-1,k-2n-2m}}R^{k-2n-m+1,m+1}
\nonumber\\
=~&R^{k-2n-m,m}-R^{k-2n-(m-1),m-1}-R^{k-2(n-1)-(m+1),m+1}=0.
\end{align}
Therefore, Eq.~\eqref{supeq:xxxR} is proved, and the recursive way to obtain $R^{k-2n-m, m}$ becomes more simple.

We note that our way to fix the degrees of freedom of $Q_{k}$ is not convenient for the case of the XXX chain because $R^{k-2n,0}=R^{k-2(n-1)-1,1}$, which corresponds to Eq.~\eqref{supeq:xxxR} for $m=0$, is not satisfied in general. However, by considering a linear combination of $Q_{k} $'s, we can obtain a set of the conserved quantities $\tilde{Q}_{k}$ 's:
\begin{gather}
\tilde{Q}_{k}=\sum_{\substack{0\leq n \leq \lfloor \frac{k}{2} \rfloor-1,\\
0\leq m \leq \lfloor \frac{k-2n}{2}\rfloor-1}}\sum_{\substack{\overline{\bm{A}}:\\
(k-2n-m,m)~\text{operators}}}\tilde{q}^{k-2n-m,m}_{\overline{A_{1}^{1+m_{1}}A_{2}^{1+m_{2}}\cdots A_{k-2n-2m-1}^{1+m_{k-2n-2m-1}}}}\ \overline{A_{1}^{1+m_{1}}A_{2}^{1+m_{2}}\cdots A_{k-2n-2m-1}^{1+m_{k-2n-2m-1}}},
\\
\tilde{q}^{k-2n-m,m}_{\overline{A_{1}^{1+m_{1}}A_{2}^{1+m_{2}}\cdots A_{k-2n-2m-1}^{1+m_{k-2n-2m-1}}}}
= s\paren{A_{1}A_{2}\cdots A_{k-2n-2m-1}}\tilde{R}^{k-2n-m, m},
\end{gather}
where $\tilde{R}^{k-2n-m, m}$ satisfies
\begin{gather}
\tilde{R}^{k-m,m}=1 \quad \text{for}~m\geq 0,
\label{supeq:tilder1}\\
\tilde{R}^{k-2n,0}=\tilde{R}^{k-2(n-1)-1,1},
\label{supeq:tilder2}\\
\tilde{R}^{k-2n-m,m}=\tilde{R}^{k-2n-(m-1),m-1}+\tilde{R}^{k-2(n-1)-(m+1),m+1} \quad \text{for}~m\geq 1.
\label{supeq:tilder3}
\end{gather}
The solution of Eqs.~\eqref{supeq:tilder1}-\eqref{supeq:tilder3} is obtained as 
\begin{align}
\tilde{R}^{k-2n-m,m}=\frac{(m+1)(2n+m)!}{n!(n+m+1)!},
\end{align}
and we have reproduced the known structure called a Catalan tree in 
Refs.~\cite{supgrabowski1994quantum,supGRABOWSKI1995299} from our procedure.
Here, $\tilde{R}^{k-2n,0}=\frac{(2n)!}{n!(n+1)!}$ is known as a Catalan number. For example, $\tilde{Q}_{10}=Q_{10}-6Q_{8}+2Q_{6}-2Q_{4}$ as shown in Fig.~\ref{supfig:tildeq}.


\end{document}